\documentclass[fleqn,usenatbib]{mnras}

\usepackage{newtxtext,newtxmath}

\usepackage[T1]{fontenc}
\usepackage{ulem}

\DeclareRobustCommand{\VAN}[3]{#2}
\let\VANthebibliography\thebibliography
\def\thebibliography{\DeclareRobustCommand{\VAN}[3]{##3}\VANthebibliography}

\usepackage{graphicx}	
\usepackage{amsmath}	
\usepackage{caption}
\usepackage{subcaption}

\usepackage{multirow}
\usepackage{array}

\usepackage{booktabs,threeparttable}

\usepackage{xfrac}

\usepackage{tikz}
\usetikzlibrary{arrows}
\usepackage[outline]{contour}
\contournumber{64}
\contourlength{.1em}
\tikzset{close/.style={outer sep=-1.5pt}}

\newcommand{\msun}{\mbox{M$_{\odot}$}}

\newcommand{\kms}{\mbox{\,$\rm{km}\,s^{-1}$}}

\newcommand{\tardis}{$\textsc{tardis}$}
\newcommand{\carsus}{$\textsc{carsus}$}
\newcommand{\tausob}{$\tau_{\mathrm{\textsc{s}}}$}
\newcommand{\ergs}{erg\,s$^{\mathrm{-1}}$}

\newcommand{\PtI}{Pt\,{\sc i}}
\newcommand{\PtII}{Pt\,{\sc ii}}
\newcommand{\PtIII}{Pt\,{\sc iii}}
\newcommand{\AuI}{Au\,{\sc i}}
\newcommand{\AuII}{Au\,{\sc ii}}
\newcommand{\AuIII}{Au\,{\sc iii}}
\newcommand{\forbPtI}{[Pt\,{\sc i}]}
\newcommand{\forbPtII}{[Pt\,{\sc ii}]}
\newcommand{\forbPtIII}{[Pt\,{\sc iii}]}
\newcommand{\forbAuI}{[Au\,{\sc i}]}
\newcommand{\forbAuII}{[Au\,{\sc ii}]}
\newcommand{\forbAuIII}{[Au\,{\sc iii}]}
\newcommand{\I}{\,{\sc i}}
\newcommand{\II}{\,{\sc ii}}
\newcommand{\III}{\,{\sc iii}}
\newcommand{\IV}{\,{\sc iv}}
\newcommand{\SrII}{Sr\,{\sc ii}}

\newcommand{\grasp}{$\textsc{grasp}^{0}$}
\newcommand{\rpro}{\mbox{$r$-process}}
\newcommand{\Aval}{\mbox{$A$-values}}

\title[Pt and Au constraints on kilonova spectra]{Constraints on the presence of platinum and gold in the spectra of the kilonova AT2017gfo}

\author[J. H. Gillanders et al.]{
J. H. Gillanders$^{1}$\thanks{E-mail: jgillanders01@qub.ac.uk},
M. McCann$^{2}$
S. A. Sim$^{1}$,
S. J. Smartt$^{1}$,
C. P. Ballance$^{2}$
\\
$^{1}$Astrophysics Research Centre, School of Mathematics and Physics, Queen’s University Belfast, BT7 1NN, UK\\
$^{2}$Centre for Theoretical Atomic, Molecular and Optical Physics, School of Mathematics and Physics, Queen’s University Belfast, BT7 1NN, UK\\
}

\date{Accepted XXX. Received YYY; in original form ZZZ}

\pubyear{2020}

\begin{document}
\label{firstpage}
\pagerange{\pageref{firstpage}--\pageref{lastpage}}
\maketitle

\begin{abstract}
Binary neutron star mergers are thought to be one of the dominant sites of production for rapid neutron capture elements, including platinum and gold. Since the discovery of the binary neutron star merger GW170817, and its associated kilonova AT2017gfo, numerous works have attempted to determine the composition of its outflowing material, but they have been hampered by the lack of complete atomic data. Here, we demonstrate how inclusion of new atomic data in synthetic spectra calculations can provide insights and constraints on the production of the heaviest elements. We employ theoretical atomic data (obtained using \grasp) for neutral, singly- and doubly-ionised platinum and gold, to generate photospheric and simple nebular phase model spectra for kilonova-like ejecta properties. We make predictions for the locations of strong transitions, which could feasibly appear in the spectra of kilonovae that are rich in these species. We identify low-lying electric quadrupole and magnetic dipole transitions that may give rise to forbidden lines when the ejecta becomes optically thin. The strongest lines lie beyond 8000\,\AA, motivating high quality near-infrared spectroscopic follow-up of kilonova candidates. We compare our model spectra to the observed spectra of AT2017gfo, and conclude that no platinum or gold signatures are prominent in the ejecta. From our nebular phase modelling, we place tentative upper limits on the platinum and gold mass of $\lesssim$ a few $10^{-3}$\,\msun, and $\lesssim 10^{-2}$\,\msun, respectively. This work demonstrates how new atomic data of heavy elements can be included in radiative transfer calculations, and motivates future searches for elemental signatures.
\end{abstract}



\section{Introduction} \label{sec:Introduction}
\par
Mergers of binary neutron star (BNS) and neutron star--black hole (NSBH) systems have long been hypothesised to be an ideal location for the synthesis of the rapid neutron capture (\rpro) elements \citep[see discussion by][]{Metzger2017}. Theoretical modelling has shown that the large neutron fraction in expelled material from these mergers is sufficient to generate these heavy elements \citep{Lattimer1974, Eichler1989, Freiburghaus1999, Rosswog1999, Goriely2011, Goriely2013, Goriely2015, Perego2014, Just2015, Sekiguchi2016}. However, spectrophotometric observations are needed to confirm the validity of the models. The first kilonova (KN; the optical counterpart of a BNS merger) was detected in 2017 \citep[AT2017gfo,][]{LigoVirgo2017,Andreoni2017,Arcavi2017,Chornock2017,Coulter2017,Cowperthwaite2017,Drout2017,Evans2017,Kasliwal2017,Lipunov2017,Nicholl2017,Pian2017,Smartt2017,Tanvir2017,Troja2017,Utsumi2017,Valenti2017}. Early theoretical models \citep{Kasen2017} predicted the overall shape of the spectra, and showed that it can be readily explained by the presence of \rpro\ material, as the associated high opacities lead to a red, long-lived component.

\par
Radiative transfer simulations for the spectra of AT2017gfo have taken two approaches. The first is to attempt direct identification of species contributing to apparent absorption features in the early spectra. \cite{Smartt2017} suggested attribution of spectral features to Te\I\ and Cs\I, elements from the second \rpro\ peak. Further work by \cite{Watson2019} attributed the same absorption features to lighter \rpro\ elements, specifically Sr\II. Both works, however, rely on incomplete atomic data. Specifically, both use data from \cite{Kurucz2017} for their models. This atomic line list provides data for the lowest few ionisation stages for all elements up to the first \rpro\ peak. However, due to the difficulties involved with generating this information for heavier elements, the line lists are mostly incomplete beyond this first peak. This makes any modelling, and subsequent conclusions difficult, as the elements without complete atomic data will be excluded from consideration.

\par
The second approach is to calculate new atomic data for heavy elements, and to model the temporal evolution of the spectra in a broad sense. The models of \cite{Kasen2017} and \cite{Tanaka2018,Tanaka2020} calculate broad-band spectral energy distributions (SEDs) for low $Y_{\rm e}$ material, which can reproduce the rising near-infrared (NIR) flux observed in AT2017gfo. The main reason why these works have focused on modelling the broad spectral shapes is due to the accuracy of the atomic data used. \cite{Kasen2017} have used elements with complete and well-calibrated\footnote{Calibrated data means that the energy levels from theoretical calculations have been effectively `shifted' to match those obtained experimentally.} atomic data to represent their less well-sampled homologues (e.g. the atomic data for the elements with $Z=21-28$ have been used as `surrogate' data for the open $d$-shell \rpro\ elements with $Z=39-48$ and $Z=72-80$). This allows them to approximate the general behaviour of these heavier open $d$-shell elements, but restricts them from predicting individual features. They also use new atomic data for the elements with $Z=58-70$, but these data have not been calibrated. Therefore they cannot be used to accurately predict the locations of individual transitions, as these will be systematically offset from the true wavelengths. Similarly, the atomic data used by \cite{Tanaka2018, Tanaka2020} have not been calibrated, and so these results also cannot be used to accurately predict the locations of individual transitions either. \cite{Fontes2020} tabulated wavelength-dependent opacities for all lanthanides and one actinide (uranium), which \cite{Even2020} used to produce SEDs, but this method also does not allow identification of specific transitions.

\par
We are focused on the first of these two methods, but our progress thus far has been hampered by the availability of accurate, complete, and reliably calibrated atomic data for the heavy elements. Hence, our focus, as outlined in this paper, is to generate atomic data for the relevant important heavy elements, with a focus on reliably calibrating to any experimental data that exist. This will enable us to base our models on accurate atomic data, and allow us to make strong predictions about individual transitions of interest. To that end, the work presented here is a pilot study, focusing specifically on platinum and gold, to demonstrate the validity and usefulness of such a study.

\par
Two of the most interesting heavy elements to search for signatures of are platinum (Pt) and gold (Au). We chose to focus our efforts on these elements for the following reasons. Pt is one of the most abundant third peak \rpro\ elements predicted to be synthesised in BNS mergers \citep[][]{Bauswein2013, Goriely2013, Goriely2015}. We want to explore the presence of heavy elements from this peak, and so Pt was an obvious choice. Au is also among the more abundant elements produced in the third peak, and so this was one reason for selecting it. Another motivation for selecting Au was to explore how much variation there may be between similar elements. Pt and Au are next to each other on the periodic table, and we would like to explore how spectroscopically similar they are. From this, we can investigate the importance of having complete atomic data for all elements abundant in BNS merger synthesis calculations, as opposed to approximating the spectral properties of one element for a whole group of elements (e.g. using Pt to approximate the behaviour of all third peak \rpro\ elements).

\par
The cosmic origin of Pt and Au is currently unknown and \cite{Kobayashi2020} argue two sites of production may be needed, with a rapid injection of the elements to explain abundance patterns in metal-poor stars. The BNS channel may have a natural time delay that cannot account for the early excess in Pt and Au observed. Collapsar accretion disks have been suggested as potential production sites of these elements, and could be the dominant source of heavy elements in the early Universe, as they do not suffer from as long a time delay \citep{Siegel2019}. Additionally, magneto-rotational supernovae (MR-SNe) have been suggested to be a site of \rpro\ nucleosynthesis \citep{Winteler2012, Mosta2018}, and could contribute to the production of Pt and Au in the early Universe. However, no distinct spectroscopic signature has yet been identified to corroborate this. For further information on potential \rpro\ element production sites, see \cite{Cowan2021}.

\par
No signature of any ion of Pt or Au has been identified in the spectra of AT2017gfo, and hence our motivation for this work is to employ recently calculated, high quality atomic data to predict possible spectral features in kilonova-type expanding ejecta.

\par
We note that there have been previous reports of observations of Pt and Au lines in astrophysical objects. \cite{Ross1972} report observations of the \AuI\ 3122.8\,\AA\ line in the solar photospheric spectrum. \cite{Jaschek1970} report detections of Pt and Au lines in the spectrum of the A-type star 73 Draconis; they observe seven \PtII\ lines, the strongest of which is the \PtII\ 4645\,\AA\ transition, and four lines belonging to \AuI. There have been other observing efforts since, which have observed more Pt and Au lines in various different stars \citep[see e.g.][]{Fuhrmann1989,Adelman1994}. Also, \cite{Castelli2004} identify two Pt lines in the spectrum of HD\,175640, a narrow lined, chemically peculiar star. They identify the \PtII\ 4061.6, 4514.1\,\AA\ transitions. They also identify two Au lines in the spectra; specifically, they detected the \AuII\ 4016.7, 4052.8\,\AA\ transitions.

\par
These line detections have all been made in the photospheric spectra of the different stars under investigation. These all have surface temperatures on the order $\sim 4000 - 10000$\,K, which is comparable to the temperature of KN ejecta at early times \citep[see][where, at $0.5 - 1.5$\,d post-merger, the characteristic black-body temperature drops from $\sim 10000$ to $\sim 5000$\,K]{Shappee2017,Smartt2017}. It is therefore expected that the strongest Pt and Au lines that have been observed in these stellar spectra could feasibly also appear in the spectra of KNe, at least while the KN exhibits similar temperatures.

\par
All these observed Pt and Au features correspond to strong permitted transitions. To date, there have been no observations of forbidden Pt or Au transitions in astrophysical sources. Forbidden lines for other elements are routinely observed in the late-time spectra of SNe \citep{Jerkstrand2017-HB}. However, the lines observed correspond to elements that are abundant in the ejecta of these transients. Since typical SNe (with the exception of collapsars and MR-SNe) are not predicted to synthesise large masses of \rpro\ material, the only \rpro\ material in the ejecta will be present from the initial formation of the star. This results in abundances much lower than that required to produce any spectral features of Pt or Au, hence the lack of observations. However, the late-phase spectra of KNe may have promise for the detection of these elements, and other \rpro\ elements, since the explosions are hypothesised to synthesise significant masses of \rpro\ material. This will result in late-phase spectra dominated by emission features from \rpro\ material, and, if Pt and Au are produced in significant quantities, then their spectral signatures could be identifiable.

\par
Our work consists of two parts. First, we determine which strong features of Pt and Au are most likely to be prominent in the early, photospheric spectra of KNe. Second, we determine whether any forbidden transitions of Pt and Au could provide emission lines in the late time, nebular spectra of KNe. In both cases, we present predictions and compare to the spectra of AT2017gfo. In Section \ref{sec:Spectra}, we detail the sources of the AT2017gfo spectra we compare our models to. In Section \ref{sec:Atomic_data}, we summarise how the atomic data we use in this work was generated. Section \ref{sec:Motivation_of_parameters} contains some motivation for the values we chose for ejecta mass in our work. In Section \ref{sec:Photospheric_phase}, we outline the steps taken to produce our photospheric phase spectral models, and we highlight our main results. Section \ref{sec:Nebular_phase} details how we generated our synthetic nebular phase spectra, and also contains our results. Finally, we summarise our work in Section \ref{sec:Conclusions}.

\section{Spectra} \label{sec:Spectra}
\par
As well as providing model spectra and predictions for the features of Pt and Au, we compare our models with the one known kilonova with a spectroscopic sequence, AT2017gfo. The data we primarily use is the set of 10 X-shooter spectra originally published by \cite{Pian2017} and \cite{Smartt2017}. As part of the ENGRAVE project, all the X-shooter spectra were flux-calibrated to a compiled set of photometric measurements taken from the published values of \cite{Andreoni2017,Arcavi2017,Chornock2017,Cowperthwaite2017,Drout2017,Evans2017,Kasliwal2017,Pian2017,Smartt2017,Tanvir2017,Troja2017,Utsumi2017,Valenti2017}. This data set is publicly available, and can be accessed through the ENGRAVE webpage\footnote{\url{www.engrave-eso.org}}, along with release notes describing the calibration, extinction correction, rest-frame velocity correction, and smoothing.

\par
We supplemented this set of spectra with data from two other sources, to extend its temporal coverage to earlier times, and to improve the spectral quality (where possible). \cite{Shappee2017} present early spectra for AT2017gfo. The spectrum taken at +0.5\,d \citep[obtained 0.9\,d before the earliest X-shooter spectrum presented by][]{Pian2017, Smartt2017} is useful for demonstrating the blue, featureless shape characteristic of early KN spectra, and so we use it here to compare with our early models. We flux-calibrated this spectrum \citep[using the \textsc{sms} code; see][]{Inserra2018} to the same compiled set of photometric points as detailed above; we also corrected for extinction, and rest-frame velocity. \cite{Tanvir2017} present \textit{HST} spectra of AT2017gfo, obtained at four separate epochs. The +9.4\,d spectrum is particularly interesting as it contains a broad, emission-like feature at $\sim 14000$\,\AA, which lies in the telluric region, and so it is not observable in our X-shooter spectrum at the same epoch. We merged this \textit{HST} spectrum with the X-shooter spectrum at this epoch, replacing the pixels of X-shooter with the flux-calibrated \textit{HST} pixels, within the telluric region.

\section{Atomic Data} \label{sec:Atomic_data}
\par
For KN observations at times, $t \gtrsim 1$\,d, and ejecta temperatures, $T \lesssim 20000$\,K, the typical ionisation stages of heavy elements are neutral up to triply ionised \citep[\I--\IV;][]{Tanaka2020}. For AT2017gfo, after $\sim 0.5$\,d, the ejecta temperature dropped to $\sim 10000$\,K, and so we expect that typical ionisation stages of heavy elements in AT2017gfo are neutral up to doubly ionised (\I--\III). Hence, for this work we focused on these ionisation stages of Pt and Au.

\par
The atomic structure for these first three ion stages of platinum and gold were calculated within a Dirac-Coulomb framework, employing the General Relativistic Atomic Structure Package \citep[\grasp;][]{Dyall1989}. The orbitals were variationally determined using a multi-configurational Dirac-Fock approach for each of the six ion stages under consideration. Our goal was to accurately determine the lowest $25 - 40$ levels of each ion stage, from a much larger configuration set. Although the National Institute of Standards and Technology Atomic Spectra Database \citep[NIST ASD;][]{NIST2020} provides a more comprehensive list of energy levels for the neutral ion stages of platinum and gold, the higher ion stages have sparse, incomplete energy state listings. Furthermore, we appreciate that the Einstein $A$-coefficients for the lowest levels of each ion stage involve non-dipole transitions that scale $\propto E^{5}$, and therefore we have utilised the option with \grasp\ to adopt spectroscopically accurate energy separations before the calculation of transition matrix elements, where available. The typical differences in the calculated wavelengths and those adopted after scaling to experimentally measured energies are $1 - 13$\,per cent for the strongest permitted transitions, and $\lesssim 4$\,per cent for the strongest forbidden transitions. The typical differences between the calculated \Aval\ and those adopted after scaling are $3 - 31$\,per cent for the strongest permitted transitions, and $1 - 14$\,per cent for the strongest forbidden transitions. For example, the 10761\,\AA\ transition ($A = 20.1$\,s$^{-1}$) of \PtI\ had a calculated wavelength of 10454\,\AA\ (and $A = 21.9$\,s$^{-1}$) in \grasp.

\par
For \PtI, the first 32 energy levels were calibrated to the energies given in the NIST ASD \citep[][]{NIST2020}. For \PtII, the first 40 levels were calibrated. For \AuI, the first 37 levels were calibrated, excluding levels 21, 28, 31 and 34, which did not have identifiable counterparts. The first 21 levels of \AuII\ were calibrated. The NIST ASD has no atomic data for either \PtIII\ or \AuIII, aside from the ground level. Therefore, all energy levels in these structures are taken as calculated in \grasp. The configurations that contain calibrated energy levels in neutral and singly-ionised platinum and gold are shown in Table \ref{tab:Configuations}. \grasp\ has also been further modified to interface with \tardis\ to provide easier future integration of new atomic data sets.

\setlength{\tabcolsep}{2pt}
\begin{table}
\centering
\begin{tabular}{c|l}
\PtI   &5d$^{10}$, 5d$^{9}$6s, 5d$^{8}$6s$^{2}$, 5d$^{9}$6p, 5d$^{8}$6s6p       \\
\PtII  &5d$^{9}$, 5d$^{8}$6s, 5d$^{7}$6s$^{2}$, 5d$^{8}$6p      \\
\AuI   &5d$^{10}$6s, 5d$^{9}$6s$^{2}$, 5d$^{10}$6p, 5d$^{9}$6s6p, 5d$^{10}$7s, 5d$^{10}$7p, 5d$^{10}$6d, 5d$^{9}$6s7s       \\
\AuII  &5d$^{10}$, 5d$^{9}$6s, 5d$^{8}$6s$^{2}$, 5d$^{9}$6p     \\
\end{tabular}
\caption{Configurations that contain calibrated energies, for neutral and singly-ionised platinum and gold. Doubly-ionised platinum and gold have no calibrated levels.}
\label{tab:Configuations}
\end{table}
\setlength{\tabcolsep}{6pt}

\par
Although these atomic structure calculations provide the foundation of any plasma modelling under LTE conditions, a companion paper (McCann et al. in prep) will provide greater detail on the atomic structure calculations, as well as the electron-impact excitation of neutral gold. This excitation calculation shall be bench-marked against the spectra of ongoing gold experiments \citep{Bromley2020} and provide insight into populating mechanisms, when the observed spectra drop out of LTE into the collisional radiative regime. This will also address the incompleteness of data for the more highly charged systems, where observed and synthetic spectra may be compared to determine the identification of higher excited states. It will also provide the mechanisms by which excited states are populated, and hopefully provide temperature and density line ratios. For an in-depth discussion of the atomic data generation, see McCann et al. (in prep). Finally, we note that all wavelengths presented throughout this paper are quoted as in vacuum.

\begin{table*}
\centering
\caption{
Input parameters used to generate the various \tardis\ models presented in this work.
}
\begin{threeparttable}
\centering
\begin{tabular}{>{\centering}p{0.12\textwidth}*{3}{>{\centering}p{0.06\textwidth}}>{\centering}p{0.001\textwidth}*{2}{>{\centering}p{0.06\textwidth}}>{\centering\arraybackslash}p{0.06\textwidth}}
\hline
\hline
    &\multicolumn{3}{c}{Pt}     &     &\multicolumn{3}{c}{Au}       \\
\cline{2-4}
\cline{6-8}
    &Model 0     &Model 1      &Model 2   &     &Model 0   &Model 1       &Model 2     \\
\hline
$t_{\rm exp}$\,(days)                   &0.5    &1.4     &2.4     &   &0.5    &1.4    &2.4    \\
$T$\,(K)                                &10000  &5000    &3700    &   &10000  &5000   &3700   \\
$v_{\rm min}$\,($c$)                    &0.30   &0.25    &0.20    &   &0.30   &0.25   &0.20   \\
$v_{\rm max}$\,($c$)                    &0.35   &0.35    &0.35    &   &0.35   &0.35   &0.35   \\
$\rho_{0}$\,($10^{-11}$ g\,cm$^{-3}$)   &0.1    &0.5     &1       &   &0.1    &1      &0.5    \\
$v_{0}$\,(\kms)                         &14000  &14000   &14000   &   &14000  &14000  &14000  \\
$t_{0}$\,(days)                         &2      &2       &2       &   &2      &2      &2      \\
$\Gamma$                                &3      &3       &3       &   &3      &3      &3      \\
$M_{\rm ej}$ ($\msun$)\tnote{*}         &0.014  &0.15    &0.50    &   &0.014  &0.30  &0.25    \\
\hline
\end{tabular}
\begin{tablenotes}
      \small
      \item[*] $M_{\rm ej}$ is a derived property of our models, but is included here for reference. It represents the mass bound by the \tardis\ computational domain, and so is a lower limit for a model's ejecta mass.
\end{tablenotes}
\end{threeparttable}
\label{tab:TARDIS_model_parameters}
\end{table*}

\section{Motivation of Parameters} \label{sec:Motivation_of_parameters}
\par
To determine what ejecta mass to use in our calculations, we considered both observational and theoretical estimates. We also performed our own calculation, to estimate Pt and Au production in KNe.

\par
Observationally, the ejecta mass can be estimated from the light curve modelling of AT2017gfo. The bolometric light curve and filter band light curves have been fit with ejecta masses between $0.01 - 0.05$\,\msun\ \citep{Cowperthwaite2017,Smartt2017,Tanvir2017,Coughlin2018,Waxman2018}.

\par
The expected mass of dynamical ejecta in BNS merger simulations is in the region $10^{-4} < M_{\rm dyn} < 10^{-2}$\,\msun\ \citep{Bauswein2013,Hotokezaka2013,Sekiguchi2016,Ciolfi2017,Radice2018}, with a large fraction of neutron-rich material ($Y_{\rm e} < 0.2$), which would favour Pt and Au production. However, disk wind ejecta can potentially be more massive ($10^{-2} < M_{\rm dw} < 10^{-1}$\,\msun), slower moving \citep[$v\simeq0.1$c,][]{Wu2016,Siegel2017}, and may be composed of either low or high $Y_{\rm e}$ material. Nuclear trajectories for low $Y_{\rm e}$ regimes \citep[][]{Bauswein2013, Goriely2013, Goriely2015} indicate that Pt and Au compositions could be as high as $5 - 15$\,per cent, by mass, which would potentially mean ejecta masses of either element of $\sim 5 \times 10^{-4} - 1.5 \times 10^{-2}$\,\msun, in low $Y_{\rm e}$ ejecta.

\par
A simple calculation to determine Pt and Au production in KNe, assuming they are the sole source of these elements in the Milky Way, also provides approximate Pt and Au masses per event. The currently accepted LIGO--Virgo rate for BNS mergers in our local Universe is $R_{\textsc{bns}} = 320^{+490}_{-240}$\,Gpc$^{-3}$\,yr$^{-1}$ \citep[][]{Abbott2020_LIGO_KN_Rates}. From \cite{Abadie2010_MW_density}, the density of Milky Way equivalent galaxies (MWEG) is $\sim 1.16 \times 10^{-2}$\,Mpc$^{-3}$. From these values, we calculated a rate of BNS mergers of $2.8^{+4.2}_{-2.1} \times 10^{-5}$\,MWEG$^{-1}$\,yr$^{-1}$, which we take as the rate of BNS mergers in the Milky Way. \cite{Asplund2009} provide estimates for the abundance of elements in the Solar System. They quote number densities for Pt and Au of $3.8 \times 10^{-11}$ and $7.7 \times 10^{-12}$, respectively. Given a MW baryonic mass of $\sim 6.4 \times 10^{10}$\,\msun\ \citep[][]{McMillan2011}, and assuming that the abundances of Pt and Au throughout the Milky Way are consistent with their Solar System abundances, we estimate that there is 378\,\msun\ of Pt, and 77\,\msun\ of Au in the galaxy. Given a MW age of $\sim 10^{10}$ years, and assuming a constant rate of BNS mergers since galaxy formation, we estimate there have been $2.8^{+4.2}_{-2.1} \times 10^{5}$\,KNe to date in the Milky Way. If we assume the sole source of Pt and Au in the galaxy comes from these mergers, then we determine that each event has to eject, on average, Pt and Au masses of $M_{\rm Pt} = 0.5-5.4 \times 10^{-3}$\,\msun\ and $M_{\rm Au} = 0.1-1.1 \times 10^{-3}$\,\msun.

\par
From our simple calculation, we predict that $\sim 5$ times more Pt will be synthesised than Au, per event. Similarly, from the nuclear trajectories of \cite{Bauswein2013, Goriely2013, Goriely2015}, we expect Pt/Au ratios of $\sim$ a few. Since the uncertainties for predicted Pt and Au masses per event are large, we do not make calculations that explore the relative masses of these elements in detail, but we do note that if either element is present, the other should also have been produced, and that the abundance of Pt is expected to be higher than Au by a moderate factor.

\par
Given all of the above, we conclude that Pt and Au masses on the order of $\sim 10^{-3}$\,\msun\ are reasonable, and so we adopt this characteristic mass for all our modelling, unless otherwise stated.

\section{Photospheric phase} \label{sec:Photospheric_phase}
\par
The spectra of AT2017gfo have been interpreted in different ways, with many authors invoking two components. A low opacity, blue component and a red contribution from high opacity ejecta are physically motivated from numerical simulations of neutron star mergers. In particular, the models of \cite{Kasen2017} have been widely used to argue for two components to explain the SED of the spectra \citep[see also e.g.][]{Chornock2017,Coughlin2018}. However, the existence of two components is by no means settled, with some work arguing that the evolution of the SED and the bolometric luminosity is dominated by one component with low to moderate opacity \citep{Smartt2017,Waxman2018}. In this paper, we model the spectra of AT2017gfo using a one-component model, to evaluate the presence of Pt and Au signatures.

\par
The focus of this photospheric phase modelling is not to fully reproduce, or `fit' the observed spectra of AT2017gfo; rather, it is to place constraints on what features the ions of Pt and Au would produce if such elements were present in the ejecta. Therefore, we calculated model compositions that were composed entirely of either Pt or Au. While these compositions are not physical, they demonstrate which features of these heavy elements are potentially detectable, and illustrate what one can do with reliable atomic data.

\par
This lack of realism is somewhat necessary since the absence of extensive atomic data for the third \rpro\ peak elements prevents a calculation encompassing all expected heavy ions. For example, nucleosynthesis calculations of \rpro\ element production in KN events produce significant mass fractions of Os and Pb, in addition to Pt and Au \citep[][]{Bauswein2013, Goriely2013, Goriely2015}. The existing data for the first few ionisation states of both Os and Pb is sparse. The NIST ASD has only 135, 97, and 41 lines for Pb\,\I, \II, and \III, respectively. For Os\,\I--\III, there are 534, 38, and 1061 lines respectively, although all lines for Os\,\III\ lie in the UV, with wavelengths $< 2100$\,\AA. The lack of a complete atomic data set, spanning the optical and NIR, prevents us from making quantitative predictions for observable features for either of these elements. A fully consistent analysis for realistic compositions can only be achieved when atomic data sets for other elements have been calculated.

\subsection{Method} \label{sec:Photospheric_phase-method}
\par
To determine if any photospheric phase spectral features could be produced by Pt or Au, we generated synthetic spectra using \tardis\ \citep{tardis, tardis2}, a one-dimensional Monte Carlo radiative transfer code capable of rapidly generating synthetic spectra for explosive transients. \tardis\ has been used previously to produce KN spectra \citep{Smartt2017, Watson2019, Perego2020}. We note that these previous works did not make use of the full relativistic treatment recently implemented for \tardis\ \citep[as outlined by][]{Vogl2019}, whereas we have incorporated this feature into our modelling here. The \tardis\ code uses a photospheric approximation, where it assumes that the properties of the transient's ejecta beneath some optically thick boundary can be approximated to a black-body with a certain temperature, either selected by the user, or determined from the code based on some requested output luminosity for the model.

\par
First, we used the new data to generate an atomic data set capable of being read by \tardis. For this, we used the \carsus\ package, which extracted the level energies and statistical weights, and also the Einstein \Aval\ for all transitions from the \grasp\ output. This was then parsed into an atomic data file for use in our \tardis\ models\footnote{The atomic data set we produced for our modelling efforts is publicly available - see Data Availability.}. For the photospheric modelling, we only included permitted lines, which are electric dipole transitions (conventionally labelled E1). Electric quadrupole (E2), magnetic dipole (M1), and higher order multipole transitions are not likely to contribute substantial opacity (or emissivity) in the diffusion phase and therefore are not expected to feature in the emergent early-phase spectra.

\par
Input parameters for the \tardis\ models are given in Table \ref{tab:TARDIS_model_parameters}. We initially used similar parameters to the models presented by \cite{Smartt2017} and \cite{Watson2019}, which broadly reproduced the early SED of AT2017gfo. We used a power law density profile for all of our models, which had the general form:
\begin{equation}
\rho(v,t_{\rm exp}) = \rho_{\rm 0}\left(\frac{t_{\rm 0}}{t_{\rm exp}}\right)^{3}\left(\frac{v}{v_{\rm 0}}\right)^{-\Gamma}
\label{eqn:Density_profile_eqn}
\end{equation}
for ${v_{\rm min}} < v < {v_{\rm max}}$, where $\rho_{\rm 0}$, $t_{\rm 0}$, $v_{\rm 0}$, $\Gamma$ and ${v_{\rm max}}$ are constants. The values for these constants were chosen empirically to reproduce the general shape of the early SED of AT2017gfo, assuming it is dominated by a single black-body component. Table \ref{tab:TARDIS_model_parameters} lists these values. The choice of photospheric ejecta velocity ($v_{\rm min}$) is in line with previous works that model the early spectra of AT2017gfo \citep[$\sim 0.2 \, c$,][]{Smartt2017, Watson2019}. The time since explosion ($t_{\rm exp}$) is well constrained based on the GW detection \citep{MMApaper2017}. A value of $\Gamma = 3$ was chosen as it typically agrees with hydrodynamical calculations \citep{Hotokezaka2013, Tanaka2013}, and has been adopted by other modelling efforts for AT2017gfo \citep[][]{Watson2019}. The values for $\rho_{0}$ were chosen either to correspond to models with a fixed ejecta mass, or such that the models had pronounced and observable features. The photospheric temperatures ($T_{\rm ph}$) chosen for the models closely resemble the black-body temperatures of the early spectra of AT2017gfo, but we allowed some variation to improve the match to the observed spectra. The photospheric luminosity ($L_{\rm ph}$) for the models is then derived using:
\begin{equation}
L_{\rm ph} = 4 \pi \left(v_{\rm min} \, t_{\rm exp}\right)^{2} \sigma \, T_{\rm ph}^{4}
\label{eqn:Luminosity_eqn}
\end{equation}
where $\sigma$ is the Stefan-Boltzmann constant. We use the \tardis\ LTE treatment for ionisation, and dilute-LTE for excitation.

\par
We present two sets of photospheric calculations in this paper, which differ primarily in ejecta mass. The term `ejecta mass' in the context of our \tardis\ models refers to the mass enclosed by the \tardis\ computational domain; i.e. it represents the amount of material in the line-forming region of the model. As such, it only represents a \textit{lower limit} for the total ejecta mass of the system, as we do not consider any material that may be enclosed beneath an optically thick inner boundary.

\begin{figure*}
\centering
\includegraphics[width=\textwidth]{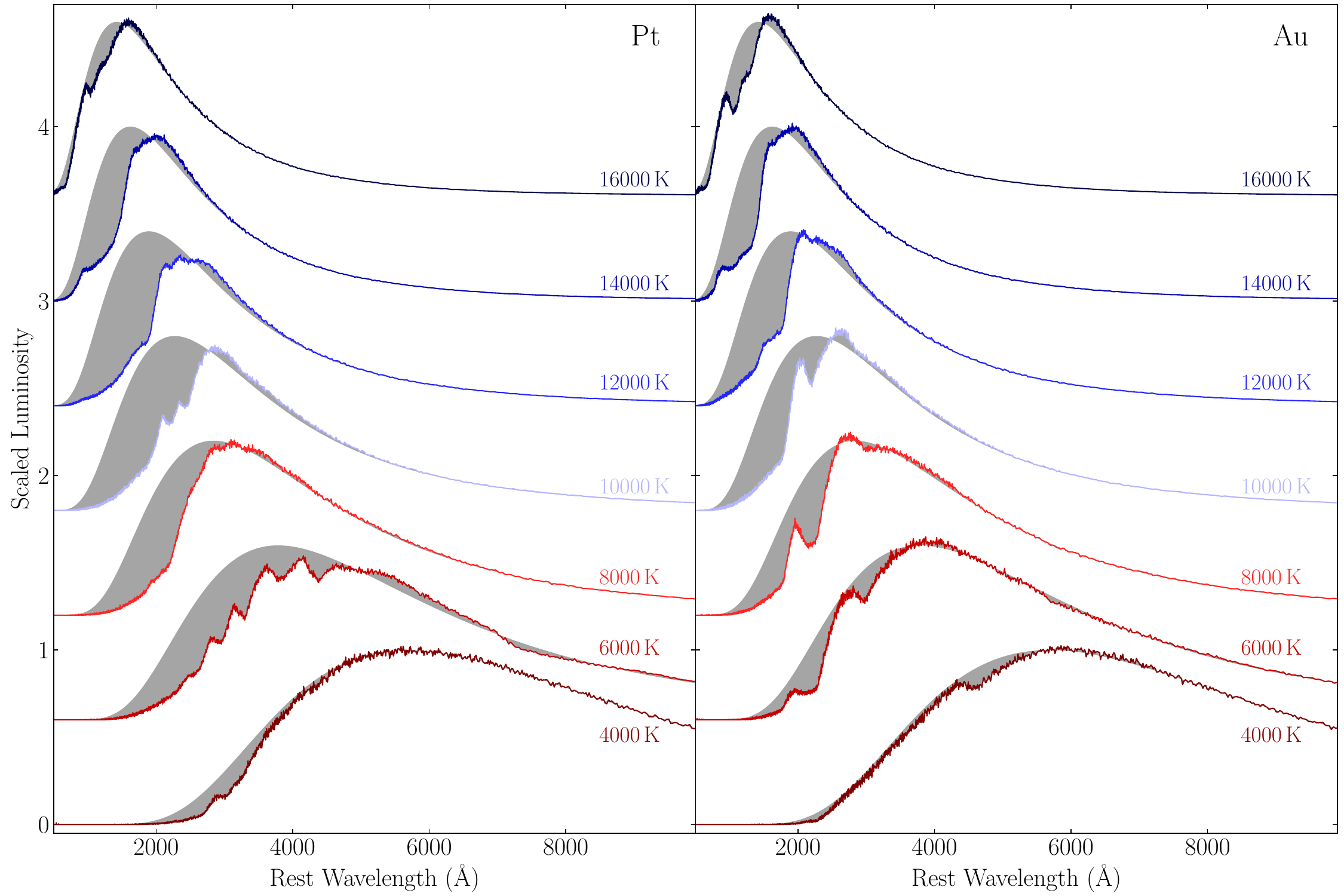}
\caption{
\tardis\ models for pure Pt and Au KNe, with varying temperatures, and $M_{\rm ej} = 10^{-3}$\,\msun. The spectra have been scaled and offset for clarity. Redward of $\sim 8000$\,\AA, there are no observable spectral features. The contribution the ejecta material has to the spectral shape is highlighted by the shaded regions. These shaded regions illustrate the deviation our model spectra have from a model with no interactions. \textit{Left panel:} Sequence of spectra for our pure Pt KN models. \textit{Right panel:} Sequence of spectra for our pure Au models.
} \label{fig:varying_T_TARDIS_model_spectra}
\end{figure*}

\begin{figure*}
\centering
\includegraphics[width=\textwidth]{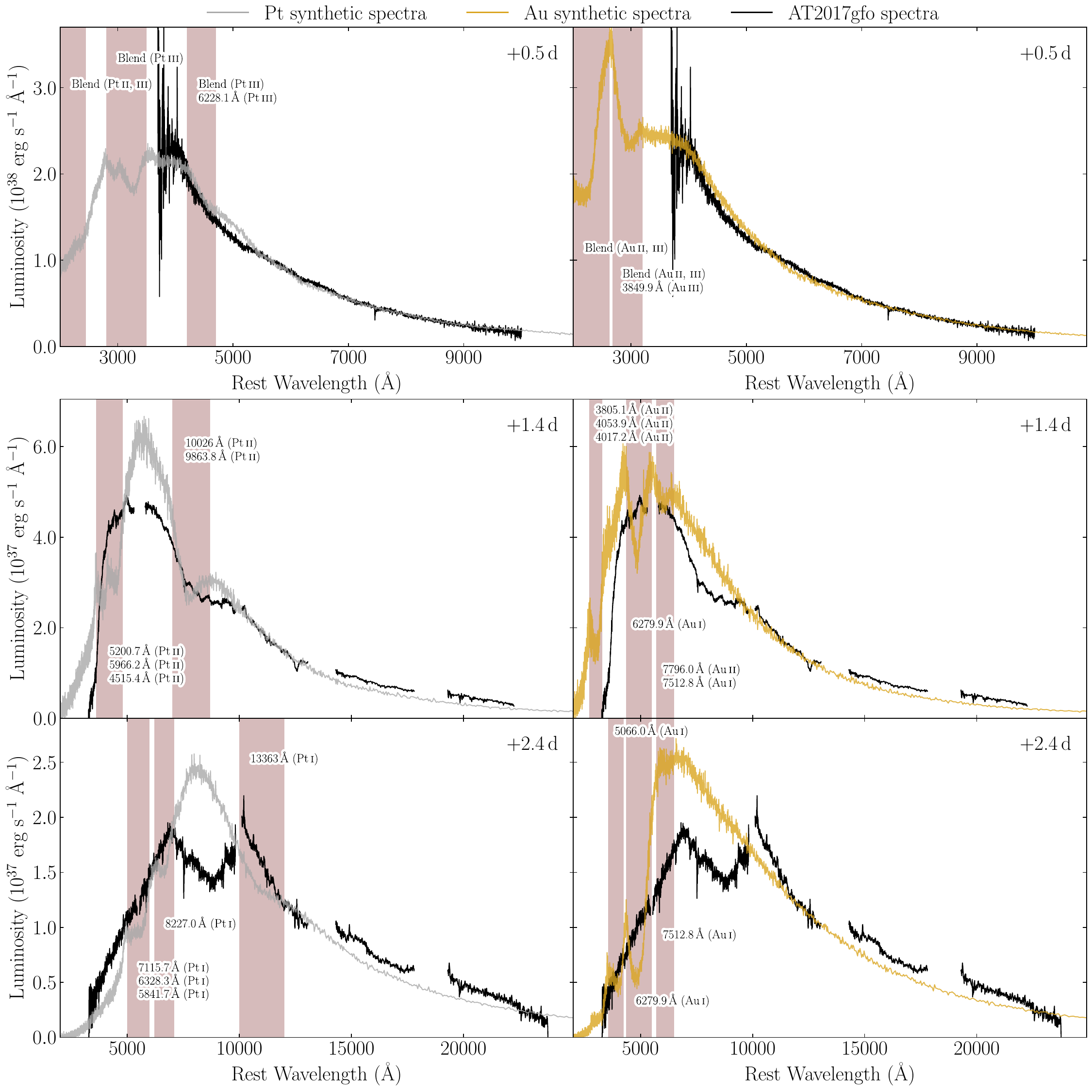}
\caption{
Comparison of our high ejecta mass \tardis\ models for pure Pt and Au compositions to early spectra of AT2017gfo. \textit{Left panels}: Model spectra for our pure Pt KN models compared to observed spectra of AT2017gfo at the corresponding epochs (+0.5, +1.4 and +2.4\,d, for the top, middle and bottom panels, respectively). \textit{Right panels}: As left, but for pure Au KN models. The regions of the spectra that deviate strongly from a black-body continuum, as a result of interaction with one (or many) strong transitions, are shaded and labelled. The shaded regions labelled with `Blend' signify that the feature is produced as a result of many different transitions, all with comparable strength. The regions where the feature was produced predominantly by a few strong transitions are labelled in descending order of contribution.
} \label{fig:TARDIS_model_spectra}
\end{figure*}

\begin{table*}
\centering
\caption{
Subset of the strongest lines in our \tardis\ models. The lines are ranked by their Sobolev optical depth (\tausob) in the models. Only lines that are highlighted in Figure \ref{fig:TARDIS_model_spectra} are included here. All lines with $\tau_{\mathrm{\textsc{s}}} > 0.01$ and $\lambda > 3000$\,\AA\ are included in the supplementary tables. The full atomic data information for each of the lower level indices indicated here can be found in Tables \ref{tab:PtI_config}--\ref{tab:AuIII_config}.
}
\begin{threeparttable}
\centering
\begin{tabular}{cccccccccc}
\hline
\hline
Species     &{\begin{tabular}[c]{@{}c@{}} \grasp\ lower \\ level index \end{tabular}}    &{\begin{tabular}[c]{@{}c@{}}  \grasp\ upper \\ level index \end{tabular}}    &{\begin{tabular}[c]{@{}c@{}} Sobolev optical \\ depth, \tausob \end{tabular}}      &{\begin{tabular}[c]{@{}c@{}} Transition \\ wavelength, $\lambda_{\rm vac}$ (\AA) \end{tabular}}      &{\begin{tabular}[c]{@{}c@{}} \grasp\ \\ $A$-value ($\mathrm{s^{-1}}$) \end{tabular}}     &{\begin{tabular}[c]{@{}c@{}} NIST ASD \\ $A$-value ($\mathrm{s^{-1}}$) \end{tabular}}       \\
\hline
\multicolumn{7}{c}{Pt model 0}      \\
\hline
\PtIII  &45\tnote{*}	  &64\tnote{*}	&0.72	&6228.1	  &3.85$\times 10^{5}$      &N/A    \\
\hline
\multicolumn{7}{c}{Au model 0}      \\
\hline
\AuIII	 &13\tnote{*}  &19\tnote{*}	&3.81	&3849.9	  &2.42$\times 10^{5}$      &N/A   \\
\hline
\multicolumn{7}{c}{Pt model 1}      \\
\hline
\PtII     &16	  &22	&12.8	 &4515.4	&5.63$\times 10^{5}$        &5.0$\times 10^{5}$	    \\
\PtII     &19	  &22	&3.61	 &5966.2	&3.24$\times 10^{5}$        &N/A	    \\
\PtII     &19	  &26	&2.71	 &5200.7	&2.93$\times 10^{5}$        &N/A	    \\
\PtII     &23	  &22	&0.85	 &10026	    &1.15$\times 10^{5}$        &N/A	    \\
\PtII     &25	  &26	&0.32	 &9863.8	&7.16$\times 10^{4}$        &N/A	    \\
\hline
\multicolumn{7}{c}{Au model 1}      \\
\hline
\AuI      &3	  &4	&2.76	 &6279.9	&1.70$\times 10^{6}$        &3.4$\times 10^{6}$	    \\
\AuII     &7	  &16	&1.44	 &4053.9	&2.08$\times 10^{6}$        &N/A       \\
\AuII     &7	  &17	&1.26	 &3805.1	&1.57$\times 10^{6}$        &N/A	    \\
\AuII     &7	  &14	&0.94	 &4017.2	&2.31$\times 10^{6}$        &N/A	    \\
\AuII     &8	  &10	&0.51	 &7796.0	&2.15$\times 10^{5}$        &N/A	    \\
\AuI      &5	  &10	&0.33	 &7512.8	&8.29$\times 10^{7}$        &4.24$\times 10^{7}$	    \\
\hline
\multicolumn{7}{c}{Pt model 2}      \\
\hline
\PtI      &9	  &15	&32.1	&5841.7	  &4.13$\times 10^{5}$      &8.0$\times 10^{5}$	    \\
\PtI      &11	  &15	&8.66	&7115.7	  &2.03$\times 10^{5}$      &9.4$\times 10^{5}$	    \\
\PtI      &12	  &24	&3.22	&6328.3	  &2.87$\times 10^{5}$      &N/A	    \\
\PtI      &12	  &18	&2.95	&8227.0	  &1.20$\times 10^{5}$      &N/A	    \\
\PtI      &13	  &18	&0.26	&13363	  &1.52$\times 10^{4}$      &N/A	    \\
\hline
\multicolumn{7}{c}{Au model 2}      \\
\hline
\AuI    	&3  &4	&27.7	  &6279.9	&1.70$\times 10^{6}$        &3.4$\times 10^{6}$	   \\
\AuI    	&3  &5	&5.62	  &5066.0   &3.27$\times 10^{5}$        &5.2$\times 10^{5}$	   \\
\AuI    	&5  &10	&0.43	  &7512.8	&8.29$\times 10^{7}$        &4.24$\times 10^{7}$	   \\
\hline
\end{tabular}
\begin{tablenotes}
      \small
      \item[*] These levels were not scaled to any experimental data. All others levels were scaled to experimentally calculated levels \citep[][]{NIST2020}.
\end{tablenotes}
\end{threeparttable}
\label{tab:TARDIS_models_strongest_lines}
\end{table*}

\subsection{Pt and Au realistic mass models} \label{sec:Photospheric_phase-Realistic_mass_models}
\par
Our first set of \tardis\ model spectra are presented in Figure \ref{fig:varying_T_TARDIS_model_spectra}. These models have ejecta masses consistent with what is expected for the Pt and Au composition in KNe. As discussed in Section \ref{sec:Motivation_of_parameters}, we chose $M_{\rm ej} = 10^{-3}$\,\msun. These models are insensitive to any material beneath the photosphere, and so they overestimate the contribution that $10^{-3}$\,\msun\ of Pt or Au would have on early KN spectra, since we have placed $10^{-3}$\,\msun\ of Pt or Au above the photosphere. This neglects the fact that some of this material is likely travelling at slower velocities, and so is hidden beneath the optically thick boundary in our model. The models cover a range of temperature, which correlates strongly with $t_{\rm exp}$. These models illustrate the effect that a modest amount of Pt and Au can have on the evolution of early KN spectra.

\subsubsection{Results} \label{sec:Photospheric_phase-Realistic_mass_models-Results}
\par
At wavelengths redward of $\sim 8000$\,\AA, where we might hope to be able to detect features due to particular transitions, we find no strong, observable spectral features. For Pt, there is strong, line blended, UV absorption ($500 - 4000$\,\AA) at each temperature (with the absorption extending into the optical for the 6000\,K model). This absorption is almost exclusively \PtIII\ absorption for the 10000\,K and hotter models, a blend of \PtII\ and \PtIII\ absorption at 8000\,K, almost exclusively \PtII\ absorption for the 6000\,K model, and almost exclusively \PtI\ for the 4000\,K model. The 16000\,K model exhibits little absorption, but this is a result of our atomic data only containing transitions up to doubly-ionised Pt. At this temperature, most of the Pt present in the ejecta is at least triply ionised, as Pt\IV. The feature at $\sim 7500$\,\AA\ in the 6000\,K model is the same feature as in the +1.4\,d high ejecta mass Pt model, presented in Figure \ref{fig:TARDIS_model_spectra}. This feature is produced by the 9863.8, 10026\,\AA\ \PtII\ transitions. Although there is visible absorption here, we require a significantly higher mass of Pt to produce a feature comparable in strength to the observed absorption feature.

\par
The Au models exhibit similar behaviour to the Pt ones. The UV absorption is almost exclusively \AuIII\ absorption for the 10000\,K and hotter models, a blend of \AuII\ and \AuIII\ absorption at 8000\,K, almost exclusively \AuII\ absorption for the 6000\,K model, and almost exclusively \AuI\ for the 4000\,K model. Similarly, the 16000\,K model exhibits little absorption, for the same reason as discussed for the Pt case. The feature at $\sim 5800$\,\AA\ in the 6000\,K model is a result of the \AuII\ 7796.0\,\AA\ transition (same feature as in the +1.4\,d high ejecta mass Au model in Figure \ref{fig:TARDIS_model_spectra}). The feature at $\sim 4500$\,\AA\ in the 4000\,K model is the same feature as in the +2.4\,d high ejecta mass Au model in Figure \ref{fig:TARDIS_model_spectra}, produced predominantly by the \AuI\ 6279.9\,\AA\ transition.

\subsubsection{Interpretation} \label{sec:Photospheric_phase-Realistic_mass_models-Interpretation}
\par
The strong UV line blanketing exhibited in these models illustrates the effect Pt and Au have on the opacity of the ejecta material, and the continuum in the UV. However, this UV absorption is not likely to be uniquely attributed to either Pt or Au in a KN spectrum, as UV line blanketing will be produced by many other heavy elements, either with $d$- or $f$-shell valence electrons. In addition, observing this region would require time-resolved spectra from a space-based telescope within the first 24\,hrs, which could only be facilitated by a rapid response of the \textit{Hubble Space Telescope}.

\par
We conclude that a mass of $\sim 10^{-3}$\,\msun\ of either Pt or Au is unlikely to produce a detectable, uniquely identifiable feature in the photospheric phase spectra of a kilonova, with the exception of the few shallow features in the optical, as discussed. These features are not well pronounced, and would require more Pt or Au mass for them to be prominent enough to be detectable.

\subsection{Pt and Au high mass models} \label{sec:Photospheric_phase-High_mass_models}
\par
Our second set of models, presented in Figure \ref{fig:TARDIS_model_spectra}, have unreasonably high masses compared to expectations for KN ejecta. The total ejecta mass for a BNS merger is likely to lie in the region $10^{-3} \lesssim M_{\rm ej} \lesssim 10^{-1}$\,\msun, which will be a mixture of \rpro\ elements (see Section \ref{sec:Motivation_of_parameters}). Our models have been constructed to determine the spectral regions showing the strongest features of Pt and Au ions that could exist in early phase KN spectra, and we find that such high mass models are necessary to illustrate these features. The masses enclosed in these \tardis\ models ($M_{\rm ej}$) are listed in Table \ref{tab:TARDIS_model_parameters}. Since we only consider pure Pt and Au models, we are neglecting the free electron contribution from other species. To quantify this effect, we generated a test \tardis\ model with equivalent Pt mass, but more total mass, such that Pt only made up 15 per cent of the total ejecta. This showed that the Pt to free electron ratio only marginally affects the resultant spectrum, and does not alter our conclusions. As discussed in Section \ref{sec:Photospheric_phase-method}, \tardis\ uses a photospheric approximation, which models optically thick material beneath some inner boundary as a black-body with a certain temperature. As a validation of our models presented here, the Planck-mean optical depths for our models \citep[calculated within \tardis\ from the sum of the Thomson scattering and line opacity, which is obtained using an expansion opacity formulation;][]{Blinnikov1998} were computed, and were confirmed to be of order unity at this inner boundary, as expected.

\par
Table \ref{tab:TARDIS_models_strongest_lines} contains the properties of a subset of the strongest permitted lines that appear in our spectra for Pt\I, \II, \III, and Au\I, \II, \III, that we have selected from our atomic data. Also included in Table \ref{tab:TARDIS_models_strongest_lines} are the \Aval\ of the transitions as quoted in the NIST ASD, where available. Some lines are completely absent from the database, and some of those that are present do not have known \Aval, hence the sparse data. From the few values that are available, there is reasonable agreement between the values predicted by \grasp, and those in the NIST ASD (mostly agree within a factor $\lesssim 2$, with the only exception being the \PtI\ 7115.7\,\AA\ transition, which varies by a factor $\lesssim 5$). While Table \ref{tab:TARDIS_models_strongest_lines} contains only the transitions that are highlighted in Figure \ref{fig:TARDIS_model_spectra}, we provide complete lists of all permitted transitions in our models with a Sobolev optical depth (calculated by \tardis), \tausob\;$ > 0.01$, and $\lambda > 3000$\,\AA. The full atomic data information for the lower levels of the transitions (electronic configuration, term, J, parity and energy) are provided in Tables \ref{tab:PtI_config}--\ref{tab:AuIII_config}.

\par
Figure \ref{fig:TARDIS_model_spectra} shows the high ejecta mass model spectra generated with \tardis, compared with three of the earliest spectra of AT2017gfo, when the photospheric regime is most likely to be applicable. The overall SED of the models approximately match the observed spectra of AT2017gfo at the same epochs, indicating that the temperatures and photospheric radii of the models are appropriate.

\subsubsection{Pt models} \label{sec:Photospheric_phase-High_mass_models-Pt_models}
\par
In the pure Pt model at +0.5\,d post-merger (Pt model 0, with parameters as in Table \ref{tab:TARDIS_model_parameters}), there is a shallow absorption feature, centred at 4500\,\AA. This is produced by a blend of \PtIII\ lines, with the most prominent being the $6228.1$\,\AA\ transition. There are two absorption features spanning the $2800 - 3500$\,\AA\ wavelength range, which are produced by a blend of many \PtIII\ lines, all of comparable strength. At wavelengths $< 2500$\,\AA, there is strong \PtII\ and \PtIII\ absorption. These strong UV features are not attributable to individual transitions; they are effectively blanket absorption due to many transitions at these wavelengths.

\par
In the second spectrum (Pt Model 1, at +1.4\,d post-merger), there is a strong absorption feature visible, with a trough at 7900\,\AA, which is due to the blue-shifted blend of the \PtII\ 9863.8, 10026\,\AA\ transitions. There are two other distinctive broad features between $3000 - 5000$\,\AA, which are the result of \PtII\ absorption, dominated by the 4515.4, 5200.7, 5966.2\,\AA\ lines.

\par
The third spectrum, Pt Model 2, calculated +2.4\,d after merger, displays a photosphere that has cooled sufficiently such that the \PtII\ lines present in the previous model have disappeared, and all observed features are now produced by \PtI. The NIR transition at 13363\,\AA\ is visible as a shallow absorption at 11000\,\AA. There are two additional absorption features, centred at 5500 and 6700\,\AA, and these are produced by the \PtI\ 5841.7, 6328.3, 7115.7 and 8227.0\,\AA\ transitions. Table \ref{tab:TARDIS_models_strongest_lines} contains the details of these strong optical and NIR lines that produce the distinct features in the \tardis\ models. 

\subsubsection{Au models} \label{sec:Photospheric_phase-High_mass_models-Au_models}
\par
In the pure Au model at +0.5\,d post-merger (Au model 0), there is strong absorption and emission due to \AuII\ and \AuIII, blueward of 3200\,\AA. The features are a result of a myriad of \AuII\ and \AuIII\ transitions, all with comparable strength. The transition with the strongest contribution in this region is the \AuIII\  3849.9\AA\ transition. No distinct transitions redward of 3000\,\AA\ are clearly visible at this temperature. 

\par
The second pure Au model (Au model 1), at +1.4\,d, has a shallow absorption feature centred at 6100\,\AA, which is produced by a blend of the \AuI\ 7512.8\,\AA\ line, and the \AuII\ 7796.0\,\AA\ line. Centred at 5000\,\AA, there is another, much more pronounced absorption feature, which is dominated by the \AuI\ 6279.9\,\AA\ line. There is also another feature at $\sim 3000$\,\AA, which is produced by the \AuII\ 3805.1, 4017.2, 4053.9\,\AA\ lines.

\par
The third pure Au model (Au model 2, at +2.4\,d post-merger) has cooled significantly from the previous epoch and all spectral features are now produced by \AuI. The absorption at $\sim 3900$\,\AA\ is dominated by the \AuI\ 5066.0\,\AA\ transition. The feature at 5000\,\AA, which was present in the previous epoch, is again produced here, predominantly by the \AuI\ 6279.9\,\AA\ transition. The absorption feature in the earlier Au model centred at $\sim 6100$\,\AA\ has become shallower, and the contribution from \AuII\ has disappeared. Now it is dominated by the \AuI\ 7512.8\,\AA\ transition.

\subsubsection{Interpretation} \label{sec:Photospheric_phase-High_mass_models-interpretation}
\par
The overall SED of the models approximately reproduce the observed spectra of AT2017gfo, as expected, since the velocities and temperatures in our \tardis\ models were chosen to produce a match to the luminosity of the transient. Although the models are simply pure Pt or Au, they do serve a purpose for determining if any absorption-like features in the photospheric spectra of AT2017gfo could be identified as Pt or Au, and for making predictions for future events.

\par
We find no plausible transition of any of the three ionisation states of Pt or Au can be uniquely matched with the features in AT2017gfo. There is strong absorption between $\sim 7000 - 10000$\,\AA\ in the observed spectra at +1.4 and +2.4\,d, which is not replicated well by contributions from Pt or Au models. Additionally, the model spectra exhibit strong absorption in regions where the observed spectra do not (e.g. Au model 1 at $\sim 5000$\,\AA). From these model spectra, it is clear that the early phase spectra of AT2017gfo are not dominated by Pt or Au features. This is not to say that there is none of these elements present in the ejecta of AT2017gfo, but it does mean that, if present, their contribution to the spectra are not prominent. It is possible that the \PtII\ blend of the 9863.8, 10026\,\AA\ transitions contribute to the broad absorption observed at 8000\,\AA, at +1.4\,d. However, this feature has been attributed to Sr\II\ by \cite{Watson2019}, and only a very large mass of Pt would produce a significant contribution.

\par
The masses of Pt and Au in the high ejecta mass models presented here represent rough estimates for upper limits of these species in the ejecta of AT2017gfo. From these, we can conclude that there is less than $\sim$ a few 0.1\,\msun\ of Pt in the ejecta. If indeed there was $\sim 0.1$\,\msun\ of Pt present, then, based on our calculation of the relative ratios of Pt and Au synthesis presented in Section \ref{sec:Motivation_of_parameters}, we would expect $\sim 0.02$\,\msun\ of Au to also be present, which we can accommodate in the early phase spectra of AT2017gfo. It should be noted however, that these masses are comparable to, or larger than, the total ejecta mass expected from a BNS merger ($10^{-3} \lesssim M_{\rm ej} \lesssim 10^{-1}$\,\msun). Therefore, our upper limits are not constraining for the amounts of Pt and Au produced by AT2017gfo.

\section{Nebular phase} \label{sec:Nebular_phase}
\par
The spectra of AT2017gfo exhibit rapid evolution, the rate of which is unprecedented, compared to other optical and NIR extragalactic transients. By $\sim 1$ week after explosion, broad features appear in the spectra that could be interpreted as being emission lines arising in a quasi-nebular plasma. Here, we make some physically motivated estimates for the different properties of a KN (temperature, density, ejecta velocity), and then make predictions for which lines we would expect to dominate the nebular phases of these transients, assuming there is some component of the ejecta made up of Pt or Au. This information is then used to generate simple, synthetic emission spectra, under the assumption of LTE level populations. Due to the simplicity of our models, we treat ionisation as a free parameter in our models, and treat the individual ions of each species independently. While such calculations are not as physically realistic as those in radiative transfer codes such as \textsc{cmfgen} \citep{Hillier-CMFGEN-1998,Dessart-CMFGEN-2005}, \textsc{artis} \citep{ARTIS-OG2009,ARTIS-neb2020}, \textsc{sumo} \citep{Jerkstrand-SUMO-2011} and \textsc{jekyll} \citep{Ergon-JEKYLL-2018}, they serve the purpose of identifying the strongest predicted transitions of these two elements.

\par
We provide a qualitative comparison with the observed spectra of AT2017gfo, in the phases where the continuum weakens and  line features become prominent. For AT2017gfo, the high ejecta velocity and low mass \citep[$v_{\rm ej} \simeq 0.2 \, c$ and $M_{\rm ej} \simeq 0.03 \, \msun$,][]{Smartt2017} imply the electron density drops to $n_{\rm e} \lesssim 10^{9}$\,cm$^{-3}$ after $\gtrsim 3$ days, assuming singly-ionised ejecta and a filling factor of 0.1, and assuming a uniform expanding sphere \citep{Jerkstrand2017-HB}. To favour radiative de-excitation, we require transitions with Einstein \Aval\;$<< 100$\,s$^{-1}$ \citep{Jerkstrand2017-HB}. Such transitions would potentially give rise to nebular emission lines in the spectra of AT2017gfo, taken after $\sim$ a few days.

\par
The spectra taken from +7.4 to +10.4\,d show broad, emission-like features at 0.79, 1.08, 1.23, 1.40, 1.58 and 2.07\,$\mu$m. In a companion paper (Gillanders et al. in prep), we propose that these are consistent with being emission features of width $35600 \pm 6600$\kms, and could be arising from optically thin, nebular-phase emission. We use this hypothesis to compare the forbidden (electric quadrupole and magnetic dipole) transitions of Pt and Au to the positions of these features. Further discussion on the nature of these late-time spectra will be provided by Gillanders et al. (in prep).

\subsection{Method} \label{sec:Nebular_phase-method}
\par
To identify candidate transitions that may appear in emission in the late-time KN spectra, we first exclude any that originate from an upper level with energy greater than the ionisation energy of the species under consideration \citep[acquired from the NIST ASD,][]{NIST2020}. Such levels would be expected to have very small populations, and therefore the corresponding transitions would have negligible contributions to our synthetic spectra.

\par
We further excluded all transitions that originate from an upper level that was not metastable, on the grounds that the upper levels of such transitions can be expected to be strongly depopulated (relative to LTE) under nebular conditions. We calculated the radiative lifetimes of all levels using:
\begin{equation} \label{eqn:radiative_lifetime}
    \tau_{\rm rad} = \left(\sum_{\textsc{L < U}} A_{\textsc{ul}} \right)^{-1}
\end{equation}
where $\tau_{\rm rad}$ is the mean radiative lifetime of the level, and $A_{\textsc{ul}}$ is the Einstein $A$-coefficient for spontaneous decay, from upper state $U$ to lower state $L$. In this work, we consider any levels with a mean radiative lifetime, $\tau_{\rm rad} \geq 10^{-2}$\,s, to be metastable (based on the discussion above). Therefore, all transitions originating from non-metastable levels (i.e. $\tau_{\rm rad} < 10^{-2}$\,s) were discarded, as the rate of emission in these transitions at late times is likely to be much lower than in LTE (which we adopt to estimate the level populations -- see below). For more on nebular phase spectra, see \cite{Jerkstrand2017-HB}. Although this leaves a large number of plausible lines (96, 593 and 1510 for \PtI, \II\ and \III\ respectively, and 10, 87 and 339 for the same three ionisation stages of Au), the vast majority of these transitions come from relatively highly excited levels, which will be heavily disfavoured in our subsequent analysis, as discussed below.

\par
To determine the approximate strengths of emission lines arising from our selected transitions, we assume LTE excitation, and estimate the population of atoms and ions in different excited states, using the Boltzmann equation:
\begin{equation} \label{eqn:no_ions_upper_lvl}
    N_{\textrm{\textsc{u}}} = N_{\textrm{\textsc{t}}} \left(\frac{g_{\textrm{\textsc{u}}}}{Z}\right) e^{-\frac{E_{\textrm{\textsc{u}}}}{k_{\textrm{\textsc{b}}} T}}
\end{equation}
where $N_{\textrm{\textsc{u}}}$ is the number of atoms or ions in the excited state, $N_{\textrm{\textsc{t}}}$ is the total number of atoms or ions, $g_{\textrm{\textsc{u}}}$ is the statistical weight of the upper level, $Z$ is the LTE partition function, $E_{\textrm{\textsc{u}}}$ is the energy of the upper level, $k_{\textrm{\textsc{b}}}$ is the Boltzmann constant, and $T$ is the temperature.

\par
With the estimates for $N_{\textrm{\textsc{u}}}$, and the Einstein \Aval\ from the atomic data, we are able to calculate the total line luminosity arising from each transition, for temperatures $T \in [2000, 3500, 5000]$\,K, assuming optically thin emission:
\begin{equation} \label{eqn:Luminosity_of_gaussians}
    L_{\rm em} = A_{\textsc{ul}} \, N_{\textrm{\textsc{u}}} \left(\frac{h c}{\lambda_{\rm vac}}\right)
\end{equation}

\par
In order to test the accuracy of our LTE assumption on the level populations, we obtained collision strength information for the lowest 5 levels of \AuI\ (see the upcoming companion paper, McCann et al. in prep). With this, we were able to calculate level populations, and compare them to our simple LTE estimates. We found that there was reasonable agreement (within a factor $\lesssim 3$), for electron densities on the order of $\sim 5 \times 10^8$\,cm$^{-3}$.

\par
The strongest transitions of  Pt\I, \II, \III\ and  Au\I, \II, \III\ in the framework of the simple LTE approximation (at a temperature of 3500\,K) are listed in Tables \ref{tab:Pt_nebular_transitions} and \ref{tab:Au_nebular_transitions}. These tables only contain the strongest transitions from our analysis; i.e. only transitions that have $L_{\rm em} > 1$\,per cent of the strongest line luminosity in our analysis are included. As expected, electric dipole transitions do not appear in either table, as these have been deselected due to the radiative lifetime cut, leaving only forbidden lines. As in Table \ref{tab:TARDIS_models_strongest_lines}, the corresponding \Aval\ for the transitions from the NIST ASD are included, where available. Again, there is good agreement between these values, and the theoretically obtained values from \grasp, with values agreeing to within a factor $\lesssim 2$. The full atomic data information for the lower and upper levels of the transitions (electronic configuration, term, J, parity and energy) are provided in Tables \ref{tab:PtI_config}--\ref{tab:AuIII_config}. We note that $\tau_{\rm sob} << 1$ for all our calculated transitions, indicating that we are in an optically thin regime, as expected.

\begin{table*}
\centering
\caption{
Strongest \forbPtI, \forbPtII\ and \forbPtIII\ transitions, at a temperature of 3500\,K. The full atomic data information for each of the level indices indicated here can be found in Tables \ref{tab:PtI_config}--\ref{tab:PtIII_config}.
}
\begin{threeparttable}
\centering
\begin{tabular}{ccccccccc}
\hline
\hline
Species   &{\begin{tabular}[c]{@{}c@{}} \grasp\ lower \\ level index \end{tabular}} &{\begin{tabular}[c]{@{}c@{}} \grasp\ upper \\ level index \end{tabular}} &{\begin{tabular}[c]{@{}c@{}} Transition \\ wavelength, $\lambda_{\rm vac}$ (\AA) \end{tabular}} &{\begin{tabular}[c]{@{}c@{}} \grasp\ \\ $A$-value ($\mathrm{s^{-1}}$) \end{tabular}}     &{\begin{tabular}[c]{@{}c@{}} NIST ASD \\ $A$-value ($\mathrm{s^{-1}}$) \end{tabular}}  &Transition type         &Relative intensity \\
\hline 
Pt\I    &2	    &6	    &10761	    &20.1	 &24.0    &M1    &1.0      \\
Pt\I    &1	    &5	    &15227	    &4.21	 &2.6     &M1    &0.45     \\
Pt\I    &3	    &7	    &10688	    &9.28	 &13.3    &M1    &0.20     \\
Pt\I    &1	    &8	    &7409.5	    &7.68	 &15.6    &M1    &0.10     \\
Pt\I    &3	    &6	    &10706	    &0.97	 &N/A     &M1    &0.049    \\
Pt\I    &3	    &9	    &6790.7	    &7.33	 &N/A     &M1    &0.045    \\
Pt\I    &5	    &9	    &11193	    &3.97	 &N/A     &M1    &0.015    \\
Pt\I    &2	    &12	    &4729.6	    &10.8	 &N/A     &M1    &0.012    \\
Pt\I    &3	    &8	    &7861.4	    &0.905	 &N/A     &M1    &0.011    \\
Pt\I    &3	    &5	    &17266	    &0.111	 &N/A     &M1    &0.011    \\
\hline 
Pt\II    &1     &3        &11877    &9.05     &8.75   &M1    &1.0          \\
Pt\II    &2     &4        &21883    &2.54     &1.38   &M1    &0.21         \\
Pt\II    &2     &8        &7512.5   &15.2     &21.1   &M1    &0.10         \\
Pt\II    &1     &6        &6332.6   &3.61     &N/A    &M1    &0.036        \\
Pt\II    &4     &7        &13397    &6.46     &13.3   &M1    &0.030        \\
Pt\II    &4     &5        &25170    &1.96     &N/A    &M1    &0.020        \\
Pt\II    &1     &4        &10688    &0.0970   &N/A    &M1    &0.016        \\
Pt\II    &1     &8        &5525.5   &1.42     &N/A    &E2    &0.013        \\
\hline 
Pt\III    &1	        &3\tnote{*}	&10917	&19.3   &N/A   &M1	  &1.0     \\
Pt\III    &2\tnote{*}	&4\tnote{*}	&12465	&8.19   &N/A   &M1    &0.026   \\
\hline
\end{tabular}
\begin{tablenotes}
      \small
      \item[*] These levels were not scaled to any experimental data. All others levels were scaled to experimentally calculated levels \citep[sourced from][]{NIST2020}.
\end{tablenotes}
\end{threeparttable}
\label{tab:Pt_nebular_transitions}
\end{table*}

\begin{table*}
\centering
\caption{
Strongest \forbAuI, \forbAuII\ and \forbAuIII\ transitions, at a temperature of 3500\,K. The full atomic data information for each of the level indices indicated here can be found in Tables \ref{tab:AuI_config}--\ref{tab:AuIII_config}. Note the \forbAuII\ 38446\,\AA\ transition, which does not appear in Figure \ref{fig:nebular_synthetic_emission_spectra}. We predict this to be one of the strongest \forbAuII\ features in our nebular phase model spectra.
}
\begin{threeparttable}
\centering
\begin{tabular}{cccccccc}
\hline
\hline
Species   &{\begin{tabular}[c]{@{}c@{}} \grasp\ lower \\ level index \end{tabular}} &{\begin{tabular}[c]{@{}c@{}} \grasp\ upper \\ level index \end{tabular}} &{\begin{tabular}[c]{@{}c@{}} Transition \\ wavelength, $\lambda_{\rm vac}$ (\AA) \end{tabular}} &{\begin{tabular}[c]{@{}c@{}} \grasp\ \\ $A$-value ($\mathrm{s^{-1}}$) \end{tabular}}     &{\begin{tabular}[c]{@{}c@{}} NIST ASD \\ $A$-value ($\mathrm{s^{-1}}$) \end{tabular}}  &Transition type     &Relative intensity \\
\hline 
Au\I    &2	    &3	    &8147.3	    &29.8	  &N/A    &M1       &1.0     \\
Au\I    &1	    &2	    &10916	    &0.0229	  &N/A    &E2       &0.13    \\
Au\I    &1	    &3	    &4665.2	    &1.20	  &N/A    &E2       &0.071   \\
\hline 
Au\II    &1   &3   &5668.7    &0.405      &N/A    &E2      &1.0     \\
Au\II    &2   &5   &6857.9    &27.1       &N/A    &M1      &0.40    \\
Au\II    &3   &4   &9876.4    &27.2       &N/A    &M1      &0.36    \\
Au\II    &1   &5   &3376.0    &8.27       &N/A    &E2      &0.25    \\
Au\II    &2   &3   &38446     &0.287      &N/A    &M1      &0.10    \\
Au\II    &3   &5   &8346.8    &1.74       &N/A    &M1      &0.021   \\
\hline 
Au\III   &1   &2\tnote{*}   &8382.3    &27.4    &N/A     &M1     &1.0        \\
\hline
\end{tabular}
\begin{tablenotes}
      \small
      \item[*] These levels were not scaled to any experimental data. All others levels were scaled to experimentally calculated levels \citep[sourced from][]{NIST2020}.
\end{tablenotes}
\end{threeparttable}
\label{tab:Au_nebular_transitions}
\end{table*}

\par
To produce a simple visualisation of how nebular emission lines may appear, we generated Gaussian emission features for all transitions. These were centred on the rest-wavelength of the transition, with a full-width, half-maximum (FWHM) velocity of 0.1\,$c$, and a peak value ($P_{\textsc{g}}$) determined from the total luminosity calculated for the transition. These peak values were determined by integrating to get the area under the Gaussian, which corresponds to the luminosity of the feature, or $L_{\rm em}$, and rearranging. This gives the expression:
\begin{equation} \label{eqn:Gaussian_height_eqn}
    P_{\textsc{g}} = 2 \sqrt{\frac{\ln(2)}{\pi}} \frac{L_{\rm em}}{\rm FWHM}
\end{equation}

\par
\begin{figure*}
\centering
\includegraphics[width=\textwidth]{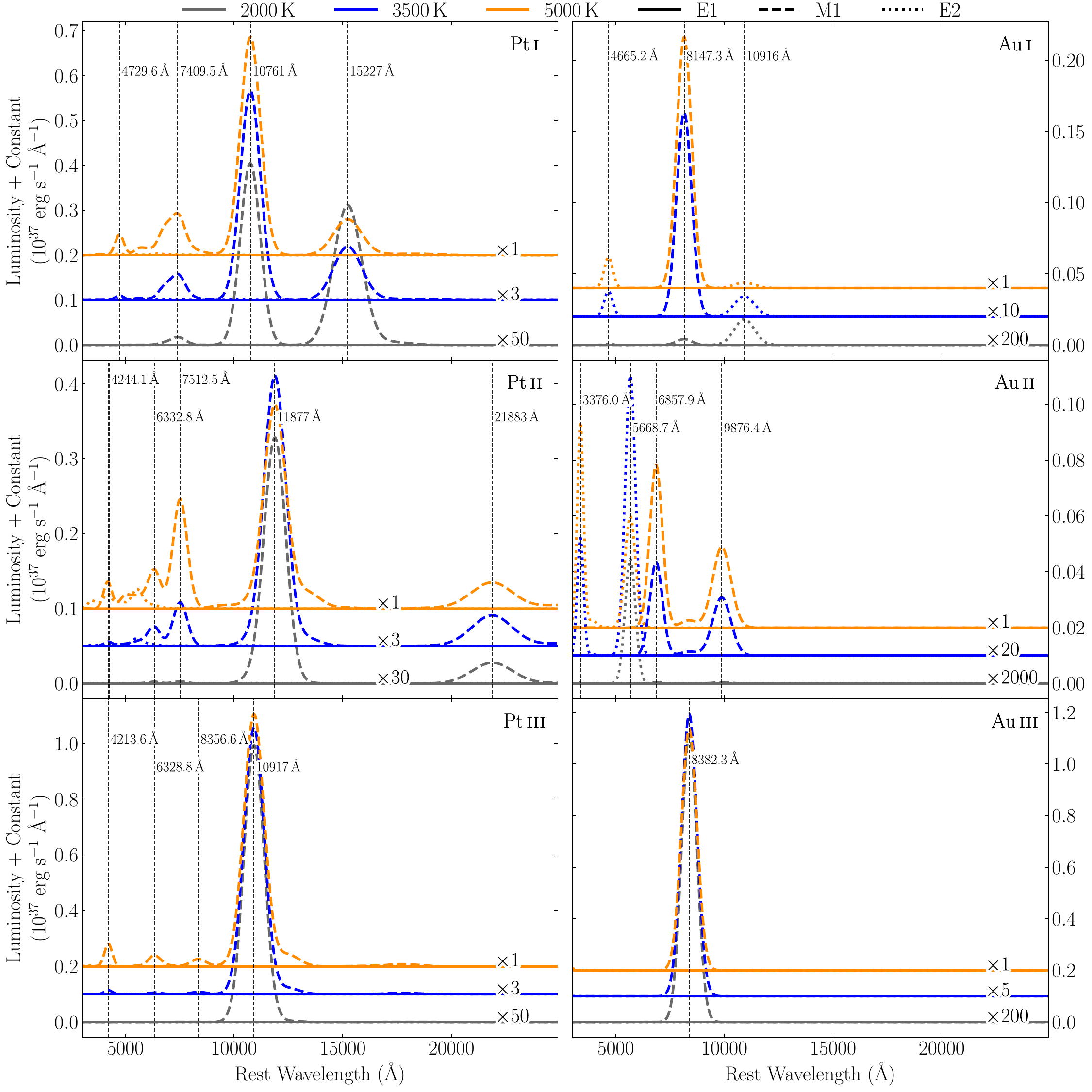}
\caption{
Synthetic emission spectra for Pt\I, \II, \III\ and Au\I, \II, \III, for a range of temperatures (2000, 3500 and 5000\,K). The emission spectra for the different transition types (E1, M1 and E2) have been plotted separately to make it clear what family the strongest emission features belong to. The spectra are offset, and scaling factors were applied, for clarity. The dominant features are labelled with their rest wavelengths, as in vacuum.
} \label{fig:nebular_synthetic_emission_spectra}
\end{figure*}

With $P_{\textsc{g}}$, we were able to compute Gaussian emission features for all transitions. Then, these Gaussians were all co-added to form one composite emission spectrum. This spectrum illustrates the relative strengths of all the transitions that we predict should be prominent in an observed nebular spectrum, which contains the species under consideration. These composite spectra are plotted in Figure \ref{fig:nebular_synthetic_emission_spectra}.

\subsection{Model results} \label{sec:Nebular_phase_model-results}
\par
Before we analyse the individual features in our models, it is worth noting that, for the Einstein \Aval\ of the strongest Pt lines, at a temperature, $T = 3500$\,K, with total ion masses, $M_{\rm ion} = 10^{-3} $\,\msun\ (which is comfortably within the range of expected Pt and Au masses discussed in Section \ref{sec:Motivation_of_parameters}), we calculate line luminosities on the order $L_{\rm Pt} \sim 10^{38}$\,\ergs. These luminosities are similar in strength to the observed features in the late-time spectra of AT2017gfo, if indeed these features are a result of emission. This motivates the study of individual features of Pt, and by extension, Au, and other third \rpro\ peak elements that are expected to be produced in KNe, as we have determined that they may be capable of producing features similar in strength to those observed in AT2017gfo.

\par
In Figure \ref{fig:nebular_synthetic_emission_spectra}, we present the LTE synthetic emission spectra for \PtI, \II, \III, and \AuI, \II, \III, at three example temperatures (2000, 3500 and 5000\,K). In each case, the ion masses are $M_{\rm ion} = 10^{-3}$\,\msun, motivated by the discussion in Section \ref{sec:Motivation_of_parameters}. The intensity of the lines scales linearly with mass, or $N_{\textrm{\textsc{t}}}$, as shown in Equations \ref{eqn:no_ions_upper_lvl} and \ref{eqn:Luminosity_of_gaussians}. We predict these lines to be the strongest features, when the ejecta has reached the optically thin regime. As discussed in Section \ref{sec:Nebular_phase-method}, low mass and high velocity ejecta can reach this regime in a few days. While high velocities and a multitude of heavy elements will likely make line blending common place in the spectra of KNe \citep{Kasen2017,Tanaka2020}, the optical and NIR wavelength range covered in Figure \ref{fig:nebular_synthetic_emission_spectra} appears to be the optimal place to observe signatures of these two elements. Tables \ref{tab:Pt_nebular_transitions} and \ref{tab:Au_nebular_transitions} indicate that these lines are all at wavelengths observable from the ground ($0.33 - 2.52 \, \mu$m), apart from the \forbAuII\ line at 3.8446\,$\mu$m.

\par
\begin{figure*}
\centering
\includegraphics[width=\textwidth]{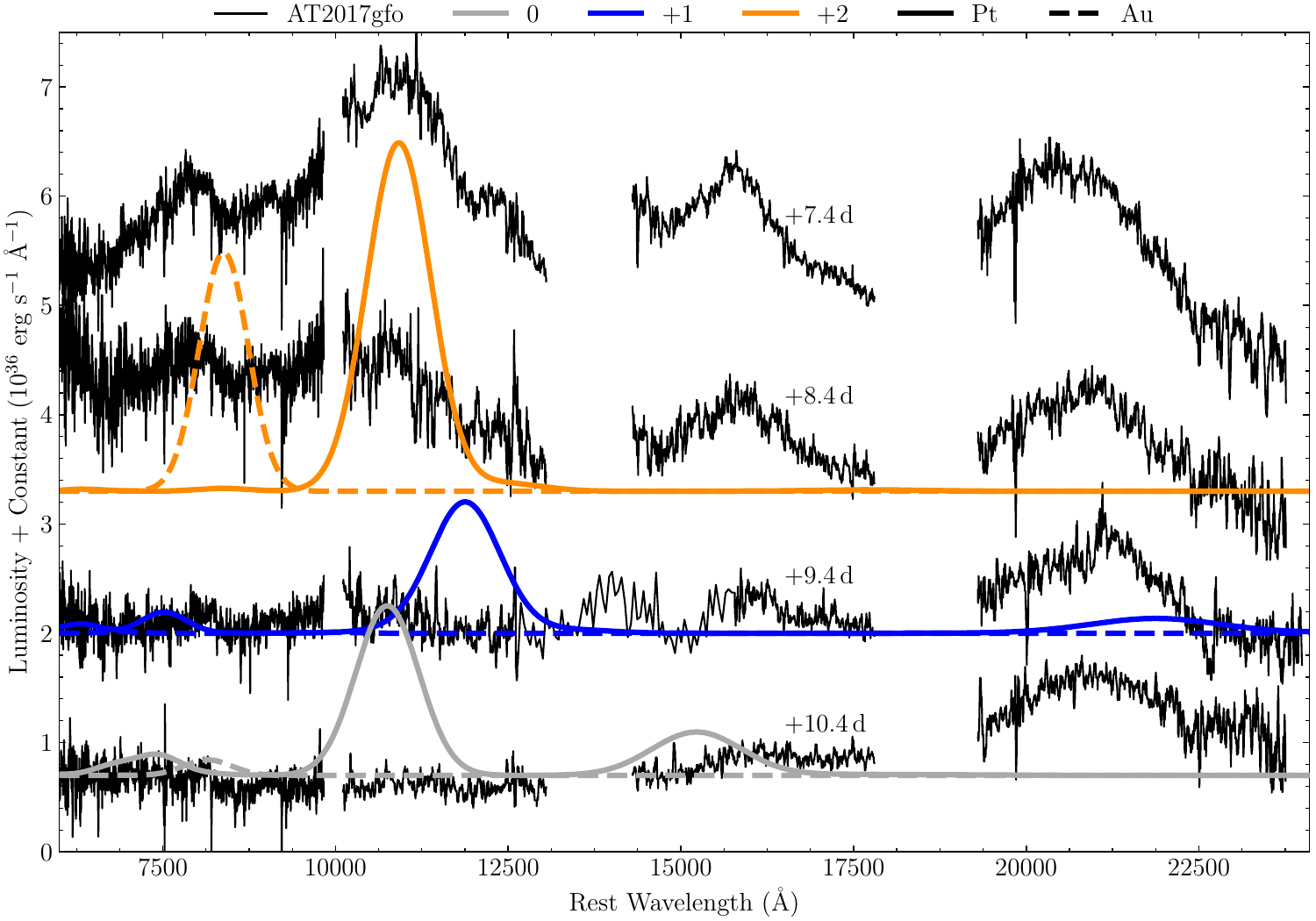}
\caption{
Comparison of the 3500\,K synthetic emission spectra for the M1 \PtI, \II, \III\ and \AuI, \II, \III\ transitions, and the late-time spectra of AT2017gfo. The spectra have been offset for clarity.
} \label{fig:comparison_of_nebular_emission_spectra_vs_17gfo}
\end{figure*}

We compare the line positions and approximate intensities to the  late-time spectra of AT2017gfo in Figure \ref{fig:comparison_of_nebular_emission_spectra_vs_17gfo}. The strongest features that could be emission lines have peaks at 0.79, 1.08, 1.23, 1.40, 1.58 and 2.07\,$\mu$m (Gillanders et al. in prep.). The strong 1.08\,$\mu$m feature weakens significantly between +7.4 and +8.4\,d. It is possible that this is the evolution of the emission component of the P-Cygni line of \SrII, identified by \cite{Watson2019}, causing the 8000\,\AA\ absorption dip in the +1.4\,d spectrum, but it could also be unrelated emission developing, coincidentally, at the same wavelength. 

\begin{figure*}
\centering
\includegraphics[width=\textwidth]{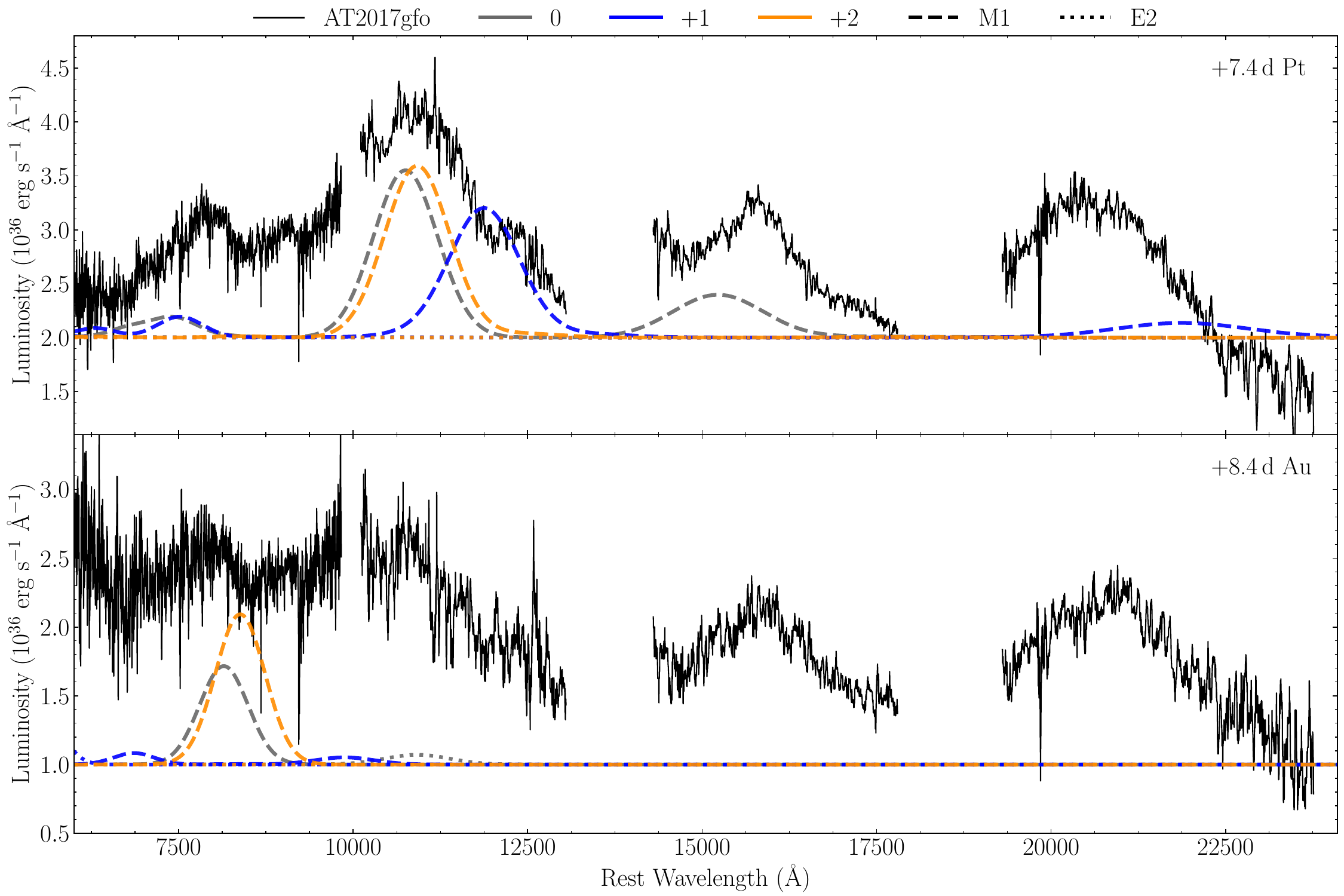}
\caption{
Comparison of the synthetic emission spectra for [Pt] and [Au] transitions, with the +7.4 and +8.4\,d AT2017gfo spectra. \textit{Upper panel:} \PtI, \II\ and \III\ M1 and E2 synthetic emission spectra plotted with the +7.4\,d AT2017gfo spectrum. We have scaled the relative ion masses arbitrarily to qualitatively match the emission features in the observed spectrum of AT2017gfo. The ion masses used for \PtI, \II, \III\ are $M_{\rm ion} = 10^{-3}, \; 10^{-3} \; \rm{and} \; 5 \times 10^{-4}$\,\msun, respectively. \textit{Lower panel:} \AuI, \II\ and \III\ M1 and E2 synthetic emission spectra plotted with the +8.4\,d AT2017gfo spectrum. Similar to the Pt synthetic emission spectra above, we have arbitrarily scaled the ion masses: $M_{\rm ion} = 5 \times 10^{-3}, \; 5 \times 10^{-3} \; \rm{and} \; 5 \times 10^{-4}$\,\msun\ for the \AuI, \II, \III\ spectra, respectively. The synthetic emission spectra are shifted to qualitatively match the continua of the observed spectra, for clarity.
} \label{fig:Pt+Au-synthetic_spectra_vs_7.4d+8.4d_gfo_spectra}
\end{figure*}

\subsubsection{Pt Models} \label{sec:Nebular_phase_model-results-Pt_models}
\par
There is an interesting coincidence between this observed emission feature, and our predicted strongest \forbPtI\ 10761\,\AA\ and \forbPtIII\ 10917\,\AA\ lines. If there was a significant contribution of \forbPtI\ 10761\,\AA\ to this feature, then the next strongest transitions of \forbPtI\ would be at 7409.5\,\AA\ and 15227\,\AA. These do not align with the peaks of the 7900\,\AA\ and  15800\,\AA\ observed features, but both of those features show asymmetric profiles, with a significant excess in the blue wing. Hence, it is possible that the 5 strongest \forbPtI\ transitions in our models are contributing emission in the +7.4\,d spectrum of AT2017gfo. Figure \ref{fig:Pt+Au-synthetic_spectra_vs_7.4d+8.4d_gfo_spectra} illustrates how different ion masses for \PtI, \II\ and \III\ could contribute to the observed features of AT2017gfo at +7.4\,d. If two (or more) of these ions were to co-exist in the ejecta at this epoch, then the spectra could be co-added to produce a composite Pt emission spectrum, which may reproduce some of the observed features.

\par
There is only one strong line of \forbPtIII, at 10917\,\AA, which lies at a similar wavelength to the strong \forbPtI\ line, at 10761\,\AA, and no further statement can be made as to the presence of this ion. The three strongest \forbPtII\ lines at 7512.5, 11877 and 21883\,\AA\ do not correspond to any of the most pronounced peaks in  the +7.4 to +10.4\,d spectra of AT2017gfo, but the 7512.5 and 11877\,\AA\ lines do lie on the asymmetric wings of observed features. We predict a \forbPtII\ line at 21883\,\AA, but this is much redder than the observed peak of the feature at 20700\,\AA. As the AT2017gfo spectra evolve from +7.4 to +10.4\,d, the emission peaks between $0.8 - 1.6 \, \mu$m weaken, and no other strong features emerge. Therefore, no further conclusive evidence for the presence of neutral or low ion stages of Pt emerge. The possible coincidences we highlight are interesting but not conclusive, and even if some \PtI\ or \PtII\ is contributing at a low or moderate level, the spectrum is dominated by other species.

\par
From the model comparisons in Figures \ref{fig:comparison_of_nebular_emission_spectra_vs_17gfo} and \ref{fig:Pt+Au-synthetic_spectra_vs_7.4d+8.4d_gfo_spectra}, we can estimate approximate upper limits for the individual ions of Pt. Clearly, $\sim 5 \times 10^{-4} - 1 \times 10^{-3}$\,\msun\ produces features comparable in strength to those observed in the late-time spectra of AT2017gfo, and so we place an upper limit of $\sim$ a few $10^{-3}$\,\msun\ on the total abundance of Pt in these spectra.

\subsubsection{Au Models} \label{sec:Nebular_phase_model-results-Au_models}
\par
In the case of Au, the two strongest transitions are close in wavelength and would be blended if they co-existed: \forbAuI\ 8147.3\,\AA, and \forbAuIII\ 8382.3\,\AA. One may dominate over the other depending on temperature and ionisation.  There is no obvious signature of an emission line or excess flux in any of the spectra of AT2017gfo at $\lambda \sim 8250$\,\AA, and any other lines of these ions would likely be weaker. The strongest \forbAuII\ lines are at 5668.7\,\AA\ and 6857.9\,\AA, but again, no obvious feature is distinguishable above the noise in the AT2017gfo spectra, although this is a region where many lines may blend together in a pseudo-continuum (Gillanders et al. in prep). Figure \ref{fig:Pt+Au-synthetic_spectra_vs_7.4d+8.4d_gfo_spectra} illustrates how different ion masses for \AuI, \II\ and \III\ may produce features comparable in strength to those observed in the +8.4\,d spectrum of AT2017gfo. It is clear that the combination of two (or more) of these ionisation stages are unable to be co-added to reproduce any of the strong observed features of AT2017gfo.

\par
From Figures \ref{fig:comparison_of_nebular_emission_spectra_vs_17gfo} and \ref{fig:Pt+Au-synthetic_spectra_vs_7.4d+8.4d_gfo_spectra}, we constrain the amount of \AuIII\ present at late times to $\sim 5 \times 10^{-4}$\,\msun. The \forbAuI\ and \forbAuII\ features are much weaker and so our models can accommodate much more of these ions ($\sim 5 \times 10^{-3}$\,\msun\ of each). From this, we place a tentative upper limit of $\sim 10^{-2}$\,\msun\ on the mass of Au present at late times, although we note that this is heavily dependent on the ionisation state of Au. From our calculations in Section \ref{sec:Motivation_of_parameters}, we predict $\sim 5$ times more Pt to be synthesised than Au. If the upper limit of Pt ($\sim$ a few $10^{-3}$\,\msun) were to be present in the ejecta, then we could easily accommodate 5 times less Au in the spectra, assuming that Pt and Au are similarly ionised.

\par
We further note that one of the strongest \forbAuII\ transitions lies beyond the observed range of the X-shooter data for AT2017gfo. The 3.8446\,$\mu$m transition would be of particular interest in the future, as the mid-infrared regime opens up with the capability of the \textit{James Webb Space Telescope} (\textit{JWST}), and the expectation that the $3 - 4 \, \mu$m region may suffer from less line-blending effects. In summary, we find no obvious coincidence with the predictions for the strongest [Au] lines and features in the spectra of AT2017gfo, at any epoch.

\section{Conclusions} \label{sec:Conclusions}
\par
The main aim of this work was to highlight the usefulness of good quality atomic data for the exploration of \rpro\ element synthesis. Here we have presented our new atomic data for neutral, singly- and doubly-ionised Pt and Au, and we have also presented some models we generated with this data. We specifically investigated the mergers of binary neutron star systems as a source of \rpro\ material in this work, and performed some spectral analysis of the kilonova AT2017gfo.

\par
First, we used \tardis\ to produce model spectra with properties similar to those expected for KNe at early times, while still in the photospheric phase (see Figures \ref{fig:varying_T_TARDIS_model_spectra} and \ref{fig:TARDIS_model_spectra}). We were able to demonstrate that, for realistic masses of Pt and Au, we see broad line-blended absorption in the UV. This property is not unique to Pt and Au, and is expected for many heavy elements. We also found that we required unrealistically large amounts of material (up to 0.5\,\msun\ in cases, see Table \ref{tab:TARDIS_model_parameters}) to produce observable individual features of any ion of Pt or Au.

\par
We generated simple emission spectra for the individual species under investigation here (\PtI, \II, \III\ and \AuI, \II, \III), using a simple LTE excitation approximation. These models are presented in Figures \ref{fig:nebular_synthetic_emission_spectra}, \ref{fig:comparison_of_nebular_emission_spectra_vs_17gfo} and \ref{fig:Pt+Au-synthetic_spectra_vs_7.4d+8.4d_gfo_spectra}. With these models, we make strong predictions for forbidden emission lines that could be detectable in the late-time, nebular-phase spectra of a KN, which has ejecta rich in these species. Many of our features lie at wavelengths $> 8000$\,\AA, demonstrating that the best method of identifying these species is through obtaining NIR spectra of future objects. X-shooter is capable of observing $\lesssim 2.5 \, \mu$m, which would capture all but one of our strongest predicted lines; the \forbAuII\ 3.8446\,$\mu$m transition would only be detectable through \textit{JWST} observations.

\par
We compared our model photospheric and nebular-phase spectra to the observed spectra of AT2017gfo. The \tardis\ model spectra were computed at the epochs of the early spectral observations of AT2017gfo (+0.5, +1.4 and +2.4\,d). We find that the strong observed feature at $\sim 7000 - 10000$\,\AA\ in the +1.4\,d spectrum may be reproducible by Pt, but the model requires a very large amount of material to do so. We identify no evidence of Au features in the early observations. We conclude that it is unlikely there are prominent Pt or Au features in the early spectra of AT2017gfo.

\par
Comparisons of our late-time model spectra with those of AT2017gfo were presented in Section \ref{sec:Nebular_phase}. At the beginning of this section, we demonstrated that, for the temperatures and masses that we have estimated for our models, the strongest forbidden transitions in our analysis have line luminosities on the order $\sim 10^{38}$\,\ergs. This demonstrates that we can expect features from these elements to be bright enough to be observed.

\par
We identify some coincidence with the \forbPtI\ 10761\,\AA\ and the \forbPtIII\ 10917\,\AA\ transitions, and a strong emission-like feature in the +7.4\,d spectrum of AT2017gfo. If there is \PtI\ present in the ejecta, then other strong features from \PtI\ should be detectable at wavelengths of 7409.5\,\AA, and 15227\,\AA. There are features near these wavelengths ($\sim 7900$\,\AA\ and $\sim 15800$\,\AA), with asymmetric profiles, displaying excess blue-wing flux. This flux could be a result of these \forbPtI\ features, and so we conclude it is plausible that there is \forbPtI\ emission in the observed late-time spectra of AT2017gfo. A definitive statement, however, will depend on future work with atomic data for many more elements, as needed to synthesise a full spectrum.

\par
It is harder to motivate the presence of \PtIII\ in the ejecta, as we only predict one prominent strong \forbPtIII\ line (at 10917\,\AA), preventing us from drawing further conclusions on its presence. We predict three strong \forbPtII\ features, none of which correspond exactly to the observed emission features, but they do lie on asymmetric wings. We conclude that it is possible that there is some contribution from Pt in the late-time spectra of AT2017gfo, but it is likely that the spectra are dominated by other species.

\par
Similar comparisons with the late-time AT2017gfo spectra and our model Au spectra do not yield such informative results. We predict a handful of strong \forbAuI, \forbAuII, and \forbAuIII\ lines, none of which correspond to observed emission features. We conclude that there is no evidence for the presence of Au in the late-time spectra of AT2017gfo.

\par
From our early, photospheric phase analysis, we were unable to meaningfully constrain the mass of Pt and Au present in the ejecta of AT2017gfo. However, our nebular phase analysis proved to be more constraining. From that, we place tentative upper limits on the Pt and Au masses of $\lesssim$ a few $10^{-3}$\,\msun, and $\lesssim 10^{-2}$\,\msun, respectively. Spectroscopic follow-up for as long as possible after any future event should be a top priority, as it is this data that we think will be the most useful for helping identify individual lines for specific elements and species in future KN events.

\par
Pt and Au are expected to be co-produced in KN ejecta. Therefore, if we observe spectral signatures for one, it is reasonable to expect to see signatures of the other. However, in Section \ref{sec:Motivation_of_parameters}, we highlighted the ratio of Pt and Au production; specifically, Pt is expected to be $\sim$ a few times more abundant than Au. Hence, it is reasonable for us to speculate that Pt may be contributing towards features in the spectra of AT2017gfo, without also detecting any Au features.

\par
Despite the fact we cannot definitively prove the presence of Pt or Au in the spectra of AT2017gfo, we have demonstrated the usefulness of having access to complete atomic data, and have also demonstrated the study that can be performed with such data. This work supports the idea that having complete atomic data for the heavy elements is useful, and we hope that these data become available in the near future. A complete set of atomic data could then be used for quantitative modelling works, where many elements are included, and more physical KN ejecta compositions are explored in detail.

\section*{Acknowledgements} \label{sec:Acknowledgements}
\par
We thank the anonymous referee for their valuable suggestions and constructive comments.
We thank Andreas Bauswein and Stephane Goriely for useful insights and discussion.
We thank Ryan Gallagher for assisting with data visualisation.
SAS, SJS, CB acknowledge funding from STFC Grants ST/P000312/1 and ST/T000198/1. 
This research made use of \tardis, a community-developed software package for spectral synthesis in supernovae.
The development of \tardis\ received support from the Google Summer of Code initiative and from ESA's Summer of Code in Space program. \tardis\ makes extensive use of Astropy and PyNE.
We are grateful for use of the computing resources from the Northern Ireland High Performance Computing (NI-HPC) service funded by EPSRC (EP/T022175).
Based on observations collected at the European Southern Observatory under ESO programmes 1102.D-0353, 0102.D-0348, 0102.D-0350, we made use of the flux-calibrated versions of the X-shooter spectra publicly available through ENGRAVE.

\section*{Data Availability} \label{sec:Data_availability}
\par
The atomic data file used for the \tardis\ modelling, and extended versions of the tables presented in this work are available, and can be accessed \href{https://pure.qub.ac.uk/en/datasets/dataset-for-the-paper-constraints-on-the-presence-of-platinum-and}{here}.

\bibliographystyle{mnras}
\bibliography{ref}

\appendix \label{sec:Appendix}
\section{Level configuration tables} \label{sec:Appendix-Level_configuration_tables}
\par
Here we present tables containing information for the relevant levels that are of interest in our work, for Pt\I, \II, \III, and Au\I, \II, \III. The tables include the level configuration and term, as well as J, parity, and energy. We have flagged any levels that were not calibrated to experimental measurements. While these tables only contain level information for the most important transitions highlighted in our work, we have included online versions, which contain all levels related to transitions presented anywhere in this work. Additionally, we have included energy level diagrams, to visualise the strongest forbidden emission transitions.

\bsp 
\newpage

\begin{table}
\centering
\caption{
\PtI\ energy levels. Only levels that are relevant to the transitions discussed in the main text have been included.
}
\begin{tabular}{cccccc}
\hline
\hline
{\begin{tabular}[c]{@{}c@{}} \grasp\ \\ level index \end{tabular}}    &Configuration   &Term   &J      &Parity       &Energy (cm$^{-1}$)    \\
\hline
1	     &5d$^{9}$6s  	            &$^{3}$D    	&3	&even	&0.00           \\
2	     &5d$^{8}$6s$^{2}$   	    &$^{3}$F    	&4	&even	&823.66         \\
3	     &5d$^{9}$6s  	            &$^{1}$D    	&2	&even	&775.88         \\
4	     &5d$^{10}$                 &$^{1}$S    	&0	&even	&6140.17        \\
5	     &5d$^{9}$6s   	            &$^{3}$D    	&2	&even	&6567.45        \\
6	     &5d$^{8}$6s$^{2}$  	    &$^{3}$F    	&3	&even	&10116.72       \\
7	     &5d$^{9}$6s   	            &$^{3}$D    	&1	&even	&10131.87       \\
8	     &5d$^{9}$6s  	            &$^{1}$D    	&2	&even	&13496.26       \\
9	     &5d$^{8}$6s$^{2}$   	    &$^{3}$F    	&2	&even	&15501.83       \\
10	     &5d$^{8}$6s$^{2}$  	    &$^{3}$P    	&0	&even	&16983.44       \\
11	     &5d$^{8}$6s$^{2}$   	    &$^{3}$P    	&1	&even	&18566.54       \\
12	     &5d$^{8}$6s$^{2}$  	    &$^{1}$G    	&4	&even	&21967.10       \\
13	     &5d$^{8}$6s$^{2}$   	    &$^{1}$D    	&2	&even	&26638.58       \\
\hline
\end{tabular}
\label{tab:PtI_config}
\end{table}

\newcommand\lvlII{0.152} 
\newcommand\lvlIII{0.0902} 
\newcommand\lvlIV{0.741} 
\newcommand\lvlV{0.834} 
\newcommand\lvlVI{1.22} 
\newcommand\lvlVII{1.29} 
\newcommand\lvlVIII{1.67}
\newcommand\lvlIX{1.92}
\newcommand\lvlX{2.11}
\newcommand\lvlXI{2.30}
\newcommand\lvlXII{2.72}
\begin{figure}
\centerline{
  \resizebox{6cm}{!}{
    \begin{tikzpicture}[
      scale=1.5,
      level/.style={thick},
      virtual/.style={thick,densely dashed},
      trans/.style={thin,->,shorten >=0.2pt,shorten <=0.2pt,>=stealth},
      classical/.style={thin,double,<->,shorten >=4pt,shorten <=4pt,>=stealth}
    ]
    \draw[level] (1cm,0em)          --      (-1cm,0em)            node[left] {\tiny1};
    \draw[level] (1cm,6em*\lvlII)   --      (-1cm,6em*\lvlII)     node[left] {\tiny2};
    \draw[level] (1cm,6em*\lvlIII)  --      (-1cm,6em*\lvlIII)    node[left] {\tiny3};
    \draw[level] (1cm,6em*\lvlIV)   --      (-1cm,6em*\lvlIV)     node[left] {\tiny4};
    \draw[level] (1cm,6em*\lvlV)    --      (-1cm,6em*\lvlV)      node[left] {\tiny5};
    \draw[level] (1cm,6em*\lvlVI)   --      (-1cm,6em*\lvlVI)     node[left] {\tiny6};
    \draw[level] (1cm,6em*\lvlVII)  --      (-1cm,6em*\lvlVII)    node[left] {\tiny7};
    \draw[level] (1cm,6em*\lvlVIII) --      (-1cm,6em*\lvlVIII)   node[left] {\tiny8};
    \draw[level] (1cm,6em*\lvlIX)   --      (-1cm,6em*\lvlIX)     node[left] {\tiny9};
    \draw[level] (1cm,6em*\lvlX)    --      (-1cm,6em*\lvlX)      node[left] {\tiny10};
    \draw[level] (1cm,6em*\lvlXI)   --      (-1cm,6em*\lvlXI)     node[left] {\tiny11};
    \draw[level] (1cm,6em*\lvlXII)  --      (-1cm,6em*\lvlXII)    node[left] {\tiny12};
    \draw[gray, trans] (-0.95cm,6em*\lvlVI)    -- (-0.95cm,6em*\lvlII)      node[midway,sloped,close,below,rotate=180] {\contour{white}{\tiny10761}};
    \draw[gray, trans] (-0.75cm,6em*\lvlV)     -- (-0.75cm,0em)             node[midway,sloped,close,below,rotate=180] {\contour{white}{\tiny15227}};
    \draw[gray, trans] (-0.55cm,6em*\lvlVII)   -- (-0.55cm,6em*\lvlIII)     node[midway,sloped,close,below,rotate=180] {\contour{white}{\tiny10688}};
    \draw[gray, trans] (-0.35cm,6em*\lvlVIII)  -- (-0.35cm,0em)             node[midway,sloped,close,below,rotate=180] {\contour{white}{\tiny7409.5}};
    \draw[gray, trans] (-0.15cm,6em*\lvlVI)    -- (-0.15cm,6em*\lvlIII)     node[midway,sloped,close,below,rotate=180] {\contour{white}{\tiny10706}};
    \draw[gray, trans] (+0.05cm,6em*\lvlIX)    -- (+0.05cm,6em*\lvlIII)     node[midway,sloped,close,below,rotate=180] {\contour{white}{\tiny6790.7}};
    \draw[gray, trans] (+0.25cm,6em*\lvlIX)    -- (+0.25cm,6em*\lvlV)       node[midway,sloped,close,below,rotate=180] {\contour{white}{\tiny11193}};
    \draw[gray, trans] (+0.45cm,6em*\lvlXII)   -- (+0.45cm,6em*\lvlII)      node[midway,sloped,close,below,rotate=180] {\contour{white}{\tiny4729.6}};
    \draw[gray, trans] (+0.65cm,6em*\lvlVIII)  -- (+0.65cm,6em*\lvlIII)     node[midway,sloped,close,below,rotate=180] {\contour{white}{\tiny7861.4}};
    \draw[gray, trans] (+0.85cm,6em*\lvlV)     -- (+0.85cm,6em*\lvlIII)     node[midway,sloped,close,below,rotate=180] {\contour{white}{\tiny17266}};
    \label{PtI energy level diagram}
    \end{tikzpicture}
  }
}
\caption{
\PtI\ energy level diagram. Only transitions that are specified in Table \ref{tab:Pt_nebular_transitions} are shown.
}
\end{figure}

\newpage

\begin{table}
\centering
\caption{
Same as Table \ref{tab:PtI_config} but for \PtII\ energy levels.
}
\begin{tabular}{cccccc}
\hline
\hline
{\begin{tabular}[c]{@{}c@{}} \grasp\ \\ level index \end{tabular}}    &Configuration   &Term   &J      &Parity       &Energy (cm$^{-1}$)    \\
\hline
1	     &5d$^{9}$ 	            &$^{2}$D   	&\sfrac{5}{2}	&even	&0.00          \\
2	     &5d$^{8}$6s 	        &$^{4}$F   	&\sfrac{9}{2}	&even	&4786.65       \\
3	     &5d$^{9}$ 	            &$^{2}$D   	&\sfrac{3}{2}	&even	&8419.84       \\
4	     &5d$^{8}$6s 	        &$^{4}$F   	&\sfrac{7}{2}	&even	&9356.32       \\
5	     &5d$^{8}$6s 	        &$^{4}$P   	&\sfrac{5}{2}	&even	&13329.28      \\
6	     &5d$^{9}$ 	            &$^{2}$D   	&\sfrac{3}{2}	&even	&15791.31      \\
7	     &5d$^{8}$6s 	        &$^{4}$F   	&\sfrac{5}{2}	&even	&16820.93      \\
8	     &5d$^{8}$6s 	        &$^{2}$F   	&\sfrac{7}{2}	&even	&18097.76      \\
16	     &5d$^{8}$6s 	        &$^{2}$G   	&\sfrac{9}{2}	&even	&29262.01      \\
19	     &5d$^{7}$6s${^2}$     	&$^{4}$F    &\sfrac{7}{2}	&even	&34647.27      \\
23	     &5d$^{7}$6s${^2}$     	&$^{4}$P    &\sfrac{5}{2}	&even	&41434.12      \\
25	     &5d$^{7}$6s${^2}$     	&$^{2}$G    &\sfrac{9}{2}	&even	&43737.43      \\
\hline
\end{tabular}
\label{tab:PtII_config}
\end{table}

\renewcommand\lvlII{0.593}
\renewcommand\lvlIII{1.04}
\renewcommand\lvlIV{1.16}
\renewcommand\lvlV{1.65}
\renewcommand\lvlVI{1.96}
\renewcommand\lvlVII{2.09}
\renewcommand\lvlVIII{2.24}
\begin{figure}
\centerline{
  \resizebox{6cm}{!}{
    \begin{tikzpicture}[
      scale=1.5,
      level/.style={thick},
      virtual/.style={thick,densely dashed},
      trans/.style={thin,->,shorten >=0.2pt,shorten <=0.2pt,>=stealth},
      classical/.style={thin,double,<->,shorten >=4pt,shorten <=4pt,>=stealth}
    ]
    \draw[level] (1cm,0em)          --  (-1cm,0em)            node[left] {\tiny1};
    \draw[level] (1cm,6em*\lvlII)   --  (-1cm,6em*\lvlII)     node[left] {\tiny2};
    \draw[level] (1cm,6em*\lvlIII)  --  (-1cm,6em*\lvlIII)    node[left] {\tiny3};
    \draw[level] (1cm,6em*\lvlIV)   --  (-1cm,6em*\lvlIV)     node[left] {\tiny4};
    \draw[level] (1cm,6em*\lvlV)    --  (-1cm,6em*\lvlV)      node[left] {\tiny5};
    \draw[level] (1cm,6em*\lvlVI)   --  (-1cm,6em*\lvlVI)     node[left] {\tiny6};
    \draw[level] (1cm,6em*\lvlVII)  --  (-1cm,6em*\lvlVII)    node[left] {\tiny7};
    \draw[level] (1cm,6em*\lvlVIII) --  (-1cm,6em*\lvlVIII)   node[left] {\tiny8};
    \draw[gray, trans] (-0.95cm,6em*\lvlIII)   -- (-0.95cm,0em)           node[midway,sloped,close,below,rotate=180] {\contour{white}{\tiny11877}};
    \draw[gray, trans] (-0.70cm,6em*\lvlIV)    -- (-0.70cm,6em*\lvlII)    node[midway,sloped,close,below,rotate=180] {\contour{white}{\tiny21883}};
    \draw[gray, trans] (-0.45cm,6em*\lvlVIII)  -- (-0.45cm,6em*\lvlII)    node[midway,sloped,close,below,rotate=180] {\contour{white}{\tiny7512.5}};
    \draw[gray, trans] (-0.20cm,6em*\lvlVI)    -- (-0.20cm,0em)           node[midway,sloped,close,below,rotate=180] {\contour{white}{\tiny6332.6}};
    \draw[gray, trans] (-0.05cm,6em*\lvlVII)   -- (-0.05cm,6em*\lvlIV)    node[midway,sloped,close,below,rotate=180] {\contour{white}{\tiny13397}};
    \draw[gray, trans] (+0.20cm,6em*\lvlV)     -- (+0.20cm,6em*\lvlIV)    node[midway,sloped,close,below,rotate=180] {\contour{white}{\tiny25170}};
    \draw[gray, trans] (+0.45cm,6em*\lvlIV)    -- (+0.45cm,0em)           node[midway,sloped,close,below,rotate=180] {\contour{white}{\tiny10688}};
    \draw[gray, trans] (+0.70cm,6em*\lvlVIII)  -- (+0.70cm,0em)           node[midway,sloped,close,below,rotate=180] {\contour{white}{\tiny5525.5}};
    \label{PtII energy level diagram}
    \end{tikzpicture}
  }
}
\caption{
\PtII\ energy level diagram. Only transitions that are specified in Table \ref{tab:Pt_nebular_transitions} are shown.
}
\vspace{0.2\textheight}
\end{figure}

\newpage

\begin{table}
\centering
\caption{
Same as Table \ref{tab:PtI_config} but for \PtIII\ energy levels.
}
\begin{threeparttable}
\centering
\begin{tabular}{cccccc}
\hline
\hline
{\begin{tabular}[c]{@{}c@{}} \grasp\ \\ level index \end{tabular}}   &Configuration   &Term   &J      &Parity       &Energy (cm$^{-1}$)    \\
\hline
1               &5d$^{8}$           &$^{3}$F    &4    &even   &0.00      \\
2\tnote{*}      &5d$^{8}$           &$^{1}$D    &2    &even   &6776.39   \\
3\tnote{*}      &5d$^{8}$           &$^{3}$F    &3    &even   &9159.88   \\
4\tnote{*}      &5d$^{8}$           &$^{3}$F    &2    &even   &14798.78  \\
45\tnote{*}     &5d$^{6}$6s$^{2}$   &$^{5}$D    &4    &even   &79582.08  \\
\hline
\end{tabular}
\begin{tablenotes}
      \small
      \item[*] These levels were not scaled to any experimental data. All others levels were scaled to experimentally calculated levels \citep[sourced from][]{NIST2020}.
\end{tablenotes}
\end{threeparttable}
\label{tab:PtIII_config}
\end{table}

\renewcommand\lvlII{0.840}
\renewcommand\lvlIII{1.14}
\renewcommand\lvlIV{1.84}
\begin{figure}
\centerline{
  \resizebox{6cm}{!}{
    \begin{tikzpicture}[
      scale=1.5,
      level/.style={thick},
      virtual/.style={thick,densely dashed},
      trans/.style={thin,->,shorten >=0.2pt,shorten <=0.2pt,>=stealth},
      classical/.style={thin,double,<->,shorten >=4pt,shorten <=4pt,>=stealth}
    ]
    \draw[level] (1cm,0em)          --      (-1cm,0em)              node[left] {\tiny1};
    \draw[level] (1cm,6em*\lvlII)   --      (-1cm,6em*\lvlII)       node[left] {\tiny2};
    \draw[level] (1cm,6em*\lvlIII)  --      (-1cm,6em*\lvlIII)      node[left] {\tiny3};
    \draw[level] (1cm,6em*\lvlIV)   --      (-1cm,6em*\lvlIV)       node[left] {\tiny4};
    \draw[gray, trans] (-0.5cm,6em*\lvlIII)   -- (-0.5cm,0em)         node[midway,sloped,close,below,rotate=180] {\contour{white}{\tiny10917}};
    \draw[gray, trans] (0.2cm,6em*\lvlIV)    -- (0.2cm,6em*\lvlII)  node[midway,sloped,close,below,rotate=180] {\contour{white}{\tiny12465}};
    \label{PtIII energy level diagram}
    \end{tikzpicture}
  }
}
\caption{
\PtIII\ energy level diagram. Only transitions that are specified in Table \ref{tab:Pt_nebular_transitions} are shown.
}
\vspace{0.3\textheight}
\end{figure}

\newpage

\begin{table}
\centering
\caption{Same as Table \ref{tab:PtI_config} but for \AuI\ energy levels.
}
\begin{tabular}{cccccc}
\hline
\hline
{\begin{tabular}[c]{@{}c@{}} \grasp\ \\ level index \end{tabular}}     &Configuration   &Term   &J      &Parity       &Energy (cm$^{-1}$)    \\
\hline
1	&5d$^{10}$6s    	  &$^{2}$S     	&$\sfrac{1}{2}$     &even	&0.00         \\
2	&5d$^{9}$6s$^{2}$ 	  &$^{2}$D     	&$\sfrac{5}{2}$ 	&even	&9161.18      \\
3	&5d$^{9}$6s$^{2}$  	  &$^{2}$D     	&$\sfrac{3}{2}$     &even	&21435.19     \\
4	&5d$^{10}$6p    	  &$^{2}$P     	&$\sfrac{1}{2}$ 	&odd	&37358.99     \\
5	&5d$^{10}$6p    	  &$^{2}$P     	&$\sfrac{3}{2}$     &odd	&41174.61     \\
\hline
\end{tabular}
\label{tab:AuI_config}
\end{table}

\renewcommand\lvlII{0.76}
\renewcommand\lvlIII{1.77}
\begin{figure}
\centerline{
  \resizebox{6cm}{!}{
    \begin{tikzpicture}[
      scale=1.5,
      level/.style={thick},
      virtual/.style={thick,densely dashed},
      trans/.style={thin,->,shorten >=0.2pt,shorten <=0.2pt,>=stealth},
      classical/.style={thin,double,<->,shorten >=4pt,shorten <=4pt,>=stealth}
    ]
    \draw[level] (1cm,0em)          --  (-1cm,0em)            node[left] {\tiny1};
    \draw[level] (1cm,6em*\lvlII)   --  (-1cm,6em*\lvlII)     node[left] {\tiny2};
    \draw[level] (1cm,6em*\lvlIII)  --  (-1cm,6em*\lvlIII)    node[left] {\tiny3};
    \draw[gray, trans] (-0.5cm,6em*\lvlIII)   -- (-0.5cm,6em*\lvlII)    node[midway,sloped,close,below,rotate=180] {\contour{white}{\tiny8147.3}};
    \draw[gray, trans] (-0.0cm,6em*\lvlII)    -- (-0.0cm,0em)           node[midway,sloped,close,below,rotate=180] {\contour{white}{\tiny10916}};
    \draw[gray, trans] (+0.5cm,6em*\lvlIII)  --  (+0.5cm,0em)           node[midway,sloped,close,below,rotate=180] {\contour{white}{\tiny4665.2}};
    \label{AuI energy level diagram}
    \end{tikzpicture}
  }
}
\caption{
\AuI\ energy level diagram. Only transitions that are specified in Table \ref{tab:Au_nebular_transitions} are shown.
}
\vspace{0.4\textheight}
\end{figure}

\newpage

\begin{table}
\centering
\caption{
Same as Table \ref{tab:PtI_config} but for \AuII\ energy levels.
}
\begin{tabular}{cccccc}
\hline
\hline
{\begin{tabular}[c]{@{}c@{}} \grasp\ \\ level index \end{tabular}}   &Configuration   &Term   &J      &Parity       &Energy (cm$^{-1}$)    \\
\hline
1	&5d$^{10}$    	        &$^{1}$S       	&0	&even	&0.00          \\
2	&5d$^{9}$6s    	        &$^{3}$D       	&3	&even	&15039.57      \\
3	&5d$^{9}$6s    	        &$^{3}$D       	&2	&even	&17640.62      \\
4	&5d$^{9}$6s    	        &$^{3}$D       	&1	&even	&27765.76      \\
5	&5d$^{9}$6s    	        &$^{1}$D       	&2	&even	&29621.25      \\
6	&5d$^{8}$6s$^{2}$   	&$^{3}$F       	&4	&even	&40478.75      \\
7	&5d$^{8}$6s$^{2}$    	&$^{1}$D       	&2	&even	&48510.89      \\ 
8	&5d$^{8}$6s$^{2}$   	&$^{3}$F       	&3	&even	&52176.51      \\
\hline
\end{tabular}
\label{tab:AuII_config}
\end{table}

\renewcommand\lvlII{0.93}
\renewcommand\lvlIII{1.095}
\renewcommand\lvlIV{1.72}
\renewcommand\lvlV{1.835}
\begin{figure}
\centerline{
  \resizebox{6cm}{!}{
    \begin{tikzpicture}[
      scale=1.5,
      level/.style={thick},
      virtual/.style={thick,densely dashed},
      trans/.style={thin,->,shorten >=0.2pt,shorten <=0.2pt,>=stealth},
      classical/.style={thin,double,<->,shorten >=4pt,shorten <=4pt,>=stealth}
    ]
    \draw[level] (1cm,0em)          --  (-1cm,0em)            node[left] {\tiny1};
    \draw[level] (1cm,6em*\lvlII)   --  (-1cm,6em*\lvlII)     node[left] {\tiny2};
    \draw[level] (1cm,6em*\lvlIII)  --  (-1cm,6em*\lvlIII)    node[left] {\tiny3};
    \draw[level] (1cm,6em*\lvlIV)   --  (-1cm,6em*\lvlIV)     node[left] {\tiny4};
    \draw[level] (1cm,6em*\lvlV)    --  (-1cm,6em*\lvlV)      node[left] {\tiny5};
    \draw[gray, trans] (-0.75cm,6em*\lvlIII)  -- (-0.75cm,0em)           node[midway,sloped,close,below,rotate=180] {\contour{white}{\tiny5668.7}};
    \draw[gray, trans] (-0.45cm,6em*\lvlV)    -- (-0.45cm,6em*\lvlII)    node[midway,sloped,close,below,rotate=180] {\contour{white}{\tiny6857.9}};
    \draw[gray, trans] (-0.15cm,6em*\lvlIV)   -- (-0.15cm,6em*\lvlIII)   node[midway,sloped,close,below,rotate=180] {\contour{white}{\tiny9876.4}};
    \draw[gray, trans] (0.15cm,6em*\lvlV)     -- (0.15cm,0em)            node[midway,sloped,close,below,rotate=180] {\contour{white}{\tiny3376.0}};
    \draw[gray, trans] (0.45cm,6em*\lvlIII)   -- (0.45cm,6em*\lvlII)     node[midway,sloped,below,rotate=180] {\contour{white}{\tiny38446}};
    \draw[gray, trans] (0.75cm,6em*\lvlV)     -- (0.75cm,6em*\lvlIII)    node[midway,sloped,close,below,rotate=180] {\contour{white}{\tiny8346.8}};
    \label{AuII energy level diagram}
    \end{tikzpicture}
  }
}
\caption{
\AuII\ energy level diagram. Only transitions that are specified in Table \ref{tab:Au_nebular_transitions} are shown.
}
\vspace{0.3\textheight}
\end{figure}

\newpage

\begin{table}
\centering
\caption{
Same as Table \ref{tab:PtI_config} but for \AuIII\ energy levels.
}
\begin{threeparttable}
\centering
\begin{tabular}{cccccc}
\hline
\hline
{\begin{tabular}[c]{@{}c@{}} \grasp\ \\ level index \end{tabular}}    &Configuration   &Term   &J      &Parity       &Energy (cm$^{-1}$)    \\
\hline
1               &5d$^{9}$      &$^{2}$D   &$\sfrac{5}{2}$   &even   &0.00        \\
2\tnote{*}      &5d$^{9}$      &$^{2}$D   &$\sfrac{3}{2}$   &even   &11929.84    \\
13\tnote{*}     &5d$^{8}$6s    &$^{2}$G   &$\sfrac{9}{2}$   &even   &59286.41    \\
\hline
\end{tabular}
\begin{tablenotes}
      \small
      \item[*] These levels were not scaled to any experimental data. All others levels were scaled to experimentally calculated levels \citep[sourced from][]{NIST2020}.
\end{tablenotes}
\end{threeparttable}
\label{tab:AuIII_config}
\end{table}

\renewcommand\lvlII{1}
\begin{figure}
\centerline{
  \resizebox{6cm}{!}{
    \begin{tikzpicture}[
      scale=1.5,
      level/.style={thick},
      virtual/.style={thick,densely dashed},
      trans/.style={thin,->,shorten >=0.2pt,shorten <=0.2pt,>=stealth},
      classical/.style={thin,double,<->,shorten >=4pt,shorten <=4pt,>=stealth}
    ]
    \draw[level] (1cm,0em)          --  (-1cm,0em)            node[left] {\tiny1};
    \draw[level] (1cm,6em*\lvlII)   --  (-1cm,6em*\lvlII)     node[left] {\tiny2};
    \draw[gray, trans] (-0.0cm,6em*\lvlII)   -- (-0.0cm,0em)    node[midway,sloped,close,below,rotate=180] {\contour{white}{\tiny8382.3}};
    \label{AuIII energy level diagram}
    \end{tikzpicture}
  }
}
\caption{
\AuIII\ energy level diagram. Only transitions that are specified in Table \ref{tab:Au_nebular_transitions} are shown.
}
\end{figure}


\label{lastpage}
\end{document}